\def\be{\begin{equation}}
    \def\ee{\end{equation}}
\def\ba{\begin{array}}
    \def\ea{\end{array}}

\documentclass[prl,showpacs,twocolumn,amsmath]{revtex4}
\usepackage{epsfig,subfigure,dsfont,amsthm,amsbsy,mathrsfs,amscd}
\usepackage{epstopdf}
\usepackage{bbm}
\usepackage{graphicx}  
\usepackage{caption}
\usepackage{braket}        
\usepackage{afterpage}        
\usepackage{float}           
\usepackage{lipsum}
\usepackage{booktabs}
\usepackage{array}
\usepackage{tabularx}

\usepackage{amsmath}
\usepackage{amsfonts}
\usepackage{mathrsfs}
\usepackage{amssymb}
\usepackage{pifont}
\usepackage{epsfig,subfigure,dsfont,amsthm,amsbsy,mathrsfs,amscd}
\def\qed{\leavevmode\unskip\penalty9999 \hbox{}\nobreak\hfill
    \quad\hbox{\leavevmode  \hbox to.77778em{%
            \hfil\vrule   \vbox to.675em%
            {\hrule width.6em\vfil\hrule}\vrule\hfil}}
    \par\vskip3pt}
\usepackage{leftidx}

\input amssym.def

\begin{document}

	\title{\large\bf Noise is not always detrimental: quantum battery capacity enhanced by the Unruh effect in noninertial frames 

}
	\author{ Xukun Wang$^{1}$, Xiaofen Huang$^{1,\dag}$, Zhihao Ma$^{2}$, Shao Ming Fei$^{3}$, and Tinggui Zhang$^{1}$}
	\affiliation{ ${1}$ School of Mathematics and Statistics, Hainan Normal University, Haikou, 571158, China \\
    ${2}$ School of Mathematical Sciences, MOE LSC, Shanghai Jiao Tong University, Shanghai 200240, China;\\
		${3}$ School of Mathematical Sciences, Capital Normal University, Beijing 100048, China \\
		$^{\dag}$ hangxf1206@163.com}

\begin{abstract}
Quantum battery capacity, as a critical metric for quantifying energy storage and release in quantum systems, exhibits complex behaviors in noninertial frames and noisy environments. This study focuses on a bipartite mixed state and aims to explore the modulation of quantum battery capacity by the Unruh effect and environmental noise. We find a counterintuitive phenomenon that the Unruh effect can enhance battery capacity, exerting a positive influence on energy storage, a result that stands in contrast to the detrimental effects usually associated with entanglement and coherence. When a quantum battery is simultaneously subjected to environmental noise and the Unruh effect, its capacity generally degrades, with the extent of degradation depending on the type of noise. The charging and discharging behaviors largely follow the same patterns observed in the noiseless scenario; however, under a bit flip channel with strong noise intensity, the charging and discharging pattern reverses. In the extreme case of maximum noise intensity, the capacity of the quantum battery under depolarizing noise tends to zero. The underlying physical mechanism lies in the fact that the bit flip channel disrupts the original population distribution of energy levels, thereby altering the average energy of the system and establishing a perturbative environment for bidirectional energy exchange. This differs fundamentally from the phase flip channel. These findings offer a new perspective for the theory of quantum batteries in noninertial reference frames.

\end{abstract}

\maketitle

\section{I. Introduction}
The pursuit of efficient energy storage devices has long been a central theme in both classical and quantum physics. Over the past decade, the concept of quantum batteries, namely quantum systems designed to store and release energy by exploiting the unique features of quantum mechanics, has emerged as a promising framework for next generation energy technologies \cite{ref8,ref1,ref2,ref3,ref4,ref5,ref6,ref7}. Unlike conventional batteries that rely on classical electrochemical processes, quantum batteries are capable of achieving superior performance through quantum correlations, coherence and collective phenomena \cite{ref9,ref10,ref11,ref12,ref13}. This concept has not only spurred deep theoretical investigations but also driven experimental explorations in quantum scale energy storage. Quantum batteries are characterized by their ability to transcend classical limitations, thereby enabling accelerated charging and discharging processes \cite{ref14,ref15,ref16,ref17,ref18,ref19,ref20,ref21,ref22,ref23,ref24,ref25}. These findings suggest that quantum batteries could be a central component of future quantum communication, computing and thermodynamic devices. 

The quantum battery capacity is tightly related to the physical ergotropy and energy level ordering \cite{ref1,ref3,ref4,ref5}. 
In Ref. \cite{refdc1}, the authors proposed a new definition for quantum battery capacity, offering a refined framework to quantify energy storage capabilities in quantum systems,
\begin{equation}\label{batt}
C(\rho, H) := \sum_{i=0}^{d-1} \epsilon_i (\lambda_{i} - \lambda_{d-1-i}),
\end{equation}
where $\rho \in \mathcal{L}(\mathcal{H})$ is a quantum state on a $d$-dimensional Hilbert space $\mathcal{H}$, $\lambda_0 \leq \lambda_1 \leq \cdots \leq \lambda_{d-1}$ are the eigenvalues of $\rho$, and $\epsilon_0 \leq \epsilon_1 \leq \cdots \leq \epsilon_{d-1}$ are the eigenenergies of the Hamiltonian $H = \sum_{i} \epsilon_i | \epsilon_i \rangle \langle \epsilon_i |$.
In the present work, a quantum battery is understood as a controllable quantum system with finite dimension whose energy storage capability is jointly specified by its state and by the battery Hamiltonian.
Moreover, Yang \emph {et al.} \cite{refdc2} experimentally validated the concept of quantum battery capacity via an optical platform and demonstrated its correlations and tradeoffs with quantum entropy, coherence and entanglement. Zhang \emph {et al.} \cite{refdc3,refdc4} proposed a scheme to enhance quantum battery capacity through local projective measurements in two qubit and three qubit systems, demonstrating capacity improvements for Bell diagonal and X shaped states. Building upon prior works, Wang \emph {et al.} \cite{refdc5} systematically analyzed the distribution relationships of quantum battery capacity, demonstrating that the sum of the capacities of two qubit X state subsystems does not exceed the total capacity. 

Significant advances in quantum battery research have highlighted their potential in idealized inertial environments. 
When quantum information tasks are considered in relativistic settings, however, the particle content of a quantum field becomes observer dependent. 
The Unruh effect states that a uniformly accelerated observer in Minkowski spacetime perceives the inertial vacuum as a thermal bath, with an effective temperature proportional to the proper acceleration \cite{refunruh1}. This observer dependent notion of particles provides a basic mechanism by which acceleration can reshape quantum correlations and information resources in noninertial frames \cite{refunruh2,refunruh3,refmann2005,refbruschi2010,refwang2010,refwangmet2020,refdong2017,refdong2024,refwclass,refwuopen2021,refwu2025jhep,refwusteer2025,reffeng2022plb,reffeng2022epl,refcoherence2022,refmi2025}.

The influence of uniform acceleration on quantum resources has been explored from several complementary perspectives. Wang \cite{refwangmet2020} \emph{et al.} showed that entangled particle detectors can improve the precision for estimating the Unruh effect. Wu \cite{refwuopen2021} \emph{et al.} investigated multipartite coherence and its monogamy relation for accelerated observers in an open system, showing that environmental decoherence and acceleration can jointly reshape coherence distribution. These representative results, together with studies of steering, uncertainty relations, multipartite entanglement, coherence and metrological probes \cite{refdong2017,refdong2024,refwclass,refwu2025jhep,refwusteer2025,reffeng2022plb,reffeng2022epl,refcoherence2022}, indicate that relativistic motion can redistribute, rather than simply destroy, quantum resources. The detailed behavior is strongly dependent on the state, partition and operational quantity under consideration.

Motivated by these developments, it is natural to ask whether the capacity of a quantum battery can also display nontrivial behavior under relativistic effects caused by acceleration. 

In this work, we consider a two cell quantum battery whose state is encoded in two Dirac field modes associated with uniformly accelerated observers. We calculate the quantum battery capacities of the accessible and inaccessible reduced states and analyze their dependence on the proper acceleration. We further include phase flip, bit flip and depolarizing channels to clarify how ordinary environmental noise competes with the redistribution of level populations caused by acceleration. Our results show that the capacity of the physically accessible battery state can increase with acceleration, whereas local noise generally suppresses the capacity; among the three noise models, bit flip noise most directly changes the population structure and can therefore produce qualitatively different charging and discharging trends. 

The remainder of this paper is organized as follows: Section II reviews the Unruh effect of Dirac fields. Section III presents the quantum battery model and its capacity under the Unruh channel. Section IV extends this analysis to noisy environments by examining the phase flip, bit flip and depolarizing channels. Section V discusses implications, limitations and future directions. Appendices provide supplementary derivations of eigenvalues and Bloch representation.

\section{ II. Unruh Effect of Dirac Fields}
We adopt natural units in which the speed of light $c$, reduced Planck constant $\hbar$ and Boltzmann constant $k_B$ are set to unity. The relativistic ingredient considered here is the Unruh effect experienced by uniformly accelerated observers in flat Minkowski spacetime. The essential point is that an inertial observer and an accelerated observer use inequivalent decompositions of the same quantum field. Therefore, the two observers do not assign the same particle content to the vacuum state.

Following the standard treatment of the Unruh effect for Dirac fields \cite{refunruh1,refunruh2,refunruh3,refwclass}, quantum fields are conveniently described by Minkowski coordinates $(t,z)$ for an inertial observer. For an observer moving with uniform proper acceleration $a$, Rindler coordinates $(\tau,\xi)$ are more suitable. In Rindler Region I, the two coordinate systems are related by
\begin{equation}
at=e^{a\xi}\sinh(a\tau),\qquad az=e^{a\xi}\cosh(a\tau),
\end{equation}
whereas in Rindler Region II they are related by
\begin{equation}
at=-e^{a\xi}\sinh(a\tau),\qquad az=-e^{a\xi}\cosh(a\tau).
\end{equation}
These transformations divide Minkowski spacetime into two causally disconnected Rindler regions. A uniformly accelerated observer in Region I has no access to the field modes in Region II. Therefore, the degrees of freedom in Region II must be traced out when one describes the state available to that observer.

Solving the Dirac equation in the Minkowski and Rindler coordinate systems gives two complete sets of field modes. The Dirac field can be expanded either in terms of Minkowski particle and antiparticle operators or in terms of Rindler particle and antiparticle operators. The corresponding Bogoliubov transformation connects the two descriptions. Under the single mode approximation, which has been widely used in studies of entanglement in noninertial frames \cite{refsinglewclass2019,refpseudopure2019,refghz2019,refwclass}, and for a fermionic mode with frequency $\omega$, this transformation can be written as
\begin{align}
a_{k,M} &= \cos r\,a_{k,\mathrm{I}}-\sin r\,b^\dagger_{k,\mathrm{II}},\\
b^\dagger_{k,M} &= \cos r\,b^\dagger_{k,\mathrm{II}}+\sin r\,a_{k,\mathrm{I}},
\end{align}
where $a_{k,M}$ and $b^\dagger_{k,M}$ are the Minkowski particle annihilation and antiparticle creation operators, while $a_{k,\mathrm{I}}$ and $b^\dagger_{k,\mathrm{II}}$ are the corresponding Rindler operators in Regions I and II.

As a consequence, the Minkowski vacuum and one particle state of a Dirac field can be expressed in terms of Rindler modes as
\begin{align}
\ket{0}_M &= \cos r \ket{0}_{\text{I}} \ket{0}_{\text{II}} + \sin r \ket{1}_{\text{I}} \ket{1}_{\text{II}}, \\
\ket{1}_M &= \ket{1}_{\text{I}} \ket{0}_{\text{II}},
\end{align}
where $\{|n\rangle_{\text{I(II)}}\}$ denote the Fock states in Rindler Region I (II). The dimensionless parameter $r$ is a convenient parametrization of the proper acceleration $a$ and the mode frequency $\omega$ through
\begin{equation}
\cos r=\frac{1}{\sqrt{e^{-2\pi\omega/a}+1}},
\end{equation}
with $r\in[0,\pi/4]$ as $a$ ranges from $0$ to $\infty$ for a fermionic field. Hence the physical variable discussed below is the proper acceleration $a$, while $r=r(a)$ is retained in the formulas and figures as its monotonic dimensionless representation.

After tracing over the inaccessible Region II modes, the transformation of a single fermionic field mode can be interpreted, at the operational level, as a quantum noise channel, often called the Unruh channel \cite{refunruh2,refwuopen2021}. The essential physical point is that the effective noise originates from the inequivalent mode decompositions used by inertial and accelerated observers, rather than from a microscopic interaction Hamiltonian between the battery and an external bath. Therefore, in the present channel description, the effect caused by acceleration is taken to modify the battery state, not the battery Hamiltonian.

\section{III. Quantum Battery Capacity under the Unruh Channel}
We consider a two cell quantum battery whose working medium is described by two effective qubits carried by field modes accessible to Alice and Bob. The initial state of the battery is chosen as the isotropic state \cite{ref40}
\begin{align} \label{iso}
\rho_{\text{iso}} &= \frac{1-p}{d^2} I \otimes I + p \ket{\psi^+} \bra{\psi^+}, 
\end{align}
where $p \in [0, 1]$ controls the weight of the maximally entangled component. For $d=2$ and $\ket{\psi^+}=\frac{1}{\sqrt2} \left(\ket{01}+\ket{10}\right)$, the isotropic state (\ref{iso}) has the Bloch representation
\begin{equation}
\rho_{AB} = \frac{1}{4} \left( I \otimes I + p \sigma_1 \otimes \sigma_1 + p \sigma_2 \otimes \sigma_2 - p \sigma_3 \otimes \sigma_3 \right), 
\end{equation}
where $\sigma_1$, $\sigma_2$, $\sigma_3$ are the standard Pauli matrices. In this setting, $\rho_{AB}$ specifies the state of the two cell battery, while the battery Hamiltonian determines the energy ordering used to evaluate its capacity.

For a spin coupled bipartite quantum battery, a general anisotropic interaction Hamiltonian may be written as \cite{refdc6,refdc7,refdc8}
\begin{align}
H=H_0+\alpha J(\sigma_1\otimes\sigma_1+\sigma_2\otimes\sigma_2)+\beta J\sigma_3\otimes\sigma_3,
\end{align}
where $H_0$ denotes the local Zeeman contribution, $J$ is the spin spin coupling strength, and $\alpha$ and $\beta$ characterize the anisotropy of the exchange interaction. In the present work we focus on the pure longitudinal Ising limit, namely $\alpha=0$, $\beta=1$, $H_0=0$ and $J=1$. With these reductions, the working Hamiltonian of the battery becomes $H = \sigma_3 \otimes \sigma_3$. This choice isolates the longitudinal two body interaction energy and makes the capacity directly sensitive to the redistribution of populations among the four joint spin configurations. Employing the definition (\ref{batt}), we have the quantum battery capacity of $\rho_{AB}$,
\begin{equation}\label{batty}
C(\rho, H) = 2 (\lambda_3 + \lambda_2 - \lambda_1 - \lambda_0), 
\end{equation}
where $\lambda_3$ and $\lambda_2$ denote the two largest eigenvalues of $\rho_{AB}$, while $\lambda_1$ and $\lambda_0$ represent the two smallest eigenvalues.

We consider Alice and Bob as observers whose battery qubits are encoded in Dirac field modes described in accelerated frames, with proper accelerations $a_A$ and $a_B$, respectively. In the formulas, these accelerations are represented by the corresponding dimensionless parameters $r_A=r(a_A)$ and $r_B=r(a_B)$. For each accelerated mode, the Region I Rindler mode is accessible, whereas the corresponding Region II mode is inaccessible. Equivalently, the initial isotropic battery state $\rho_{AB}$ is transformed by local Unruh channels into an enlarged state $\rho_{A_{\text{I}} A_{\text{II}} B_{\text{I}} B_{\text{II}}}$, and the operationally relevant battery state is obtained after tracing over the inaccessible Region II modes.
Physically, this construction should be understood as a relativistic noise channel description. The Unruh channel acts on the density operator of the field mode qubits, in the same operational sense as a quantum channel acting on a state. In this way, the degradation or redistribution of quantum information caused by acceleration is incorporated through the state transformation.
By tracing over the modes $A_{\text{II}}$ and $B_{\text{II}}$, we obtain the reduced bipartite mixed state $\rho_{A_{\text{I}}B_{\text{I}}}$,
\begin{align}
\rho_{A_{\text{I}} B_{\text{I}}} &= \frac{1}{4} \Big( I \otimes I - \sin^2 r_A \sigma_3 \otimes I - \sin^2 r_B I \otimes \sigma_3 \nonumber \\
&\quad + p \cos r_A \cos r_B \sigma_1 \otimes \sigma_1 + p \cos r_A \cos r_B \sigma_2 \otimes \sigma_2 \nonumber \\
&\quad + ( \sin^2 r_A \sin^2 r_B - p \cos^2 r_A \cos^2 r_B ) \sigma_3 \otimes \sigma_3 \Big),
\end{align}
with eigenvalues given by
\begin{align*}
\lambda_0 &= \frac{1}{4} \Big( (1 - \sin^2 r_A) (1 - \sin^2 r_B) - p \cos^2 r_A \cos^2 r_B \Big), \\
\lambda_1 &= \frac{1}{4} \Big( 1 - \sin^2 r_A \sin^2 r_B + p \cos^2 r_A \cos^2 r_B \\
&\quad - \sqrt{ (\sin^2 r_A - \sin^2 r_B)^2 + 4p^2 \cos^2 r_A \cos^2 r_B } \Big), \\
\lambda_2 &= \frac{1}{4} \Big( (1 + \sin^2 r_A) (1 + \sin^2 r_B) - p \cos^2 r_A \cos^2 r_B \Big), \\
\lambda_3 &= \frac{1}{4} \Big( 1 - \sin^2 r_A \sin^2 r_B + p \cos^2 r_A \cos^2 r_B \\
&\quad + \sqrt{ (\sin^2 r_A - \sin^2 r_B)^2 + 4p^2 \cos^2 r_A \cos^2 r_B } \Big).
\end{align*}

In an analogous manner, we compute the remaining three reduced density operators,
\begin{align}
\rho_{A_{\text{I}} B_{\text{II}}} & = \frac{1}{4} \Big( I \otimes I - \sin^2 r_A \sigma_3 \otimes I + \cos^2 r_B I \otimes \sigma_3 \nonumber \\
&\quad + p \cos r_A \sin r_B \sigma_1 \otimes \sigma_1 - p \cos r_A \sin r_B \sigma_2 \otimes \sigma_2 \nonumber \\
&\quad + ( p \cos^2 r_A \sin^2 r_B - \sin^2 r_A \cos^2 r_B ) \sigma_3 \otimes \sigma_3 \Big).  
\end{align}
\begin{align}
\rho_{A_{\text{II}} B_{\text{I}}} & = \frac{1}{4} \Big( I \otimes I + \cos^2 r_A \sigma_3 \otimes I - \sin^2 r_B I \otimes \sigma_3 \nonumber \\
&\quad + p \sin r_A \cos r_B \sigma_1 \otimes \sigma_1 - p \sin r_A \cos r_B \sigma_2 \otimes \sigma_2 \nonumber \\
&\quad + ( p \sin^2 r_A \cos^2 r_B - \cos^2 r_A \sin^2 r_B ) \sigma_3 \otimes \sigma_3 \Big).  
\end{align}
\begin{align}
\rho_{A_{\text{II}} B_{\text{II}}} & = \frac{1}{4} \Big( I \otimes I + \cos^2 r_A \sigma_3 \otimes I + \cos^2 r_B I \otimes \sigma_3 \nonumber \\
&\quad + p \sin r_A \sin r_B \sigma_1 \otimes \sigma_1 + p \sin r_A \sin r_B \sigma_2 \otimes \sigma_2 \nonumber \\
&\quad + ( \cos^2 r_A \cos^2 r_B - p \sin^2 r_A \sin^2 r_B ) \sigma_3 \otimes \sigma_3 \Big). 
\end{align}
The eigenvalues of above states are provided in the Appendix A.
 
\begin{figure*}[t]
\centering
\subfigure[]{\includegraphics[width=0.24\textwidth]{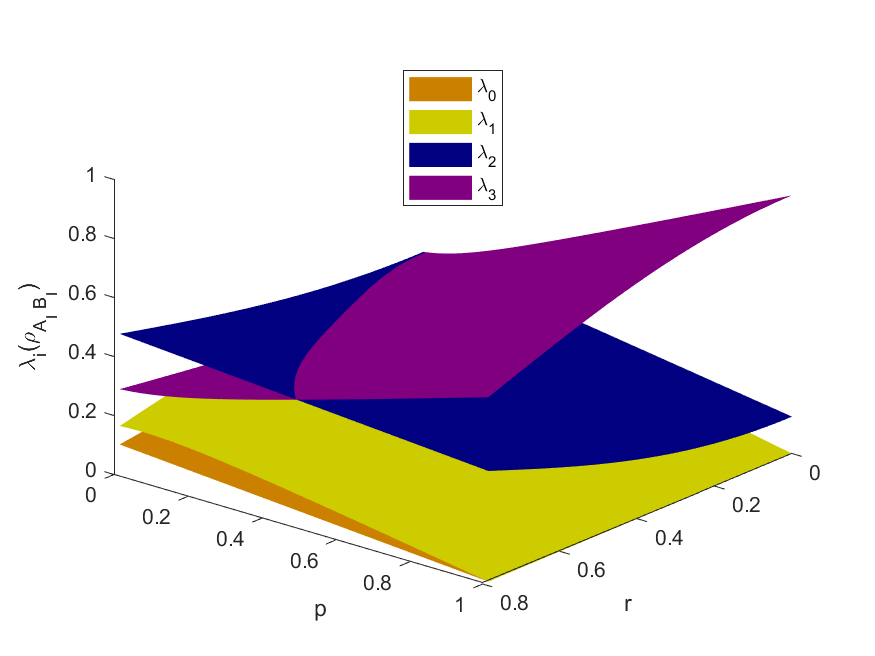}}
\subfigure[]{\includegraphics[width=0.24\textwidth]{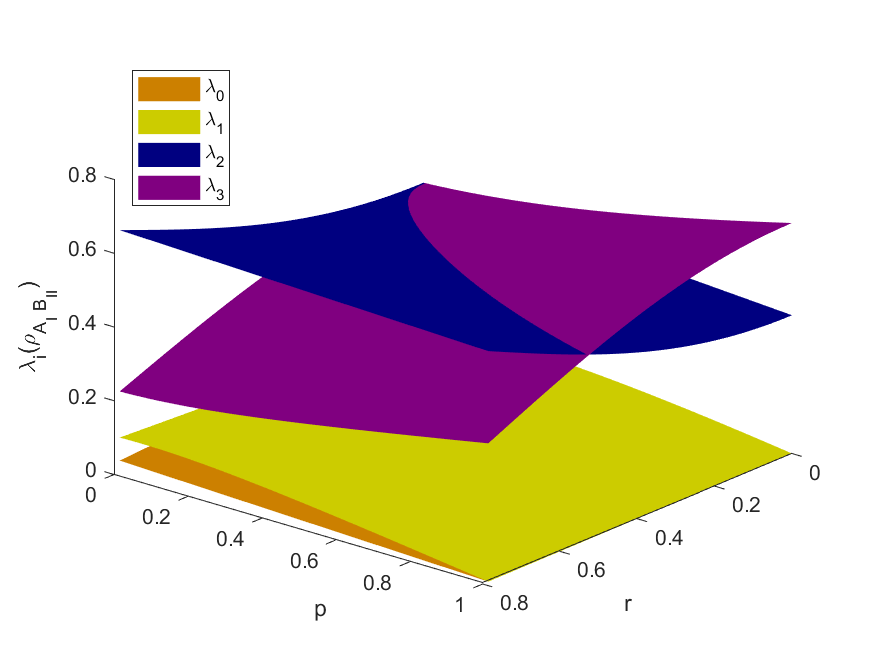}}
\subfigure[]{\includegraphics[width=0.24\textwidth]{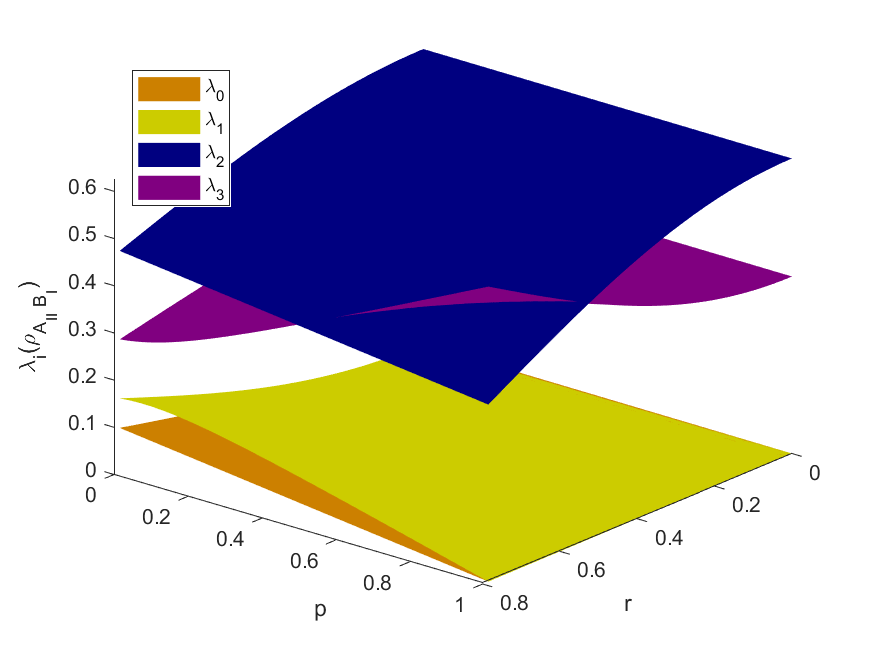}}
\subfigure[]{\includegraphics[width=0.24\textwidth]{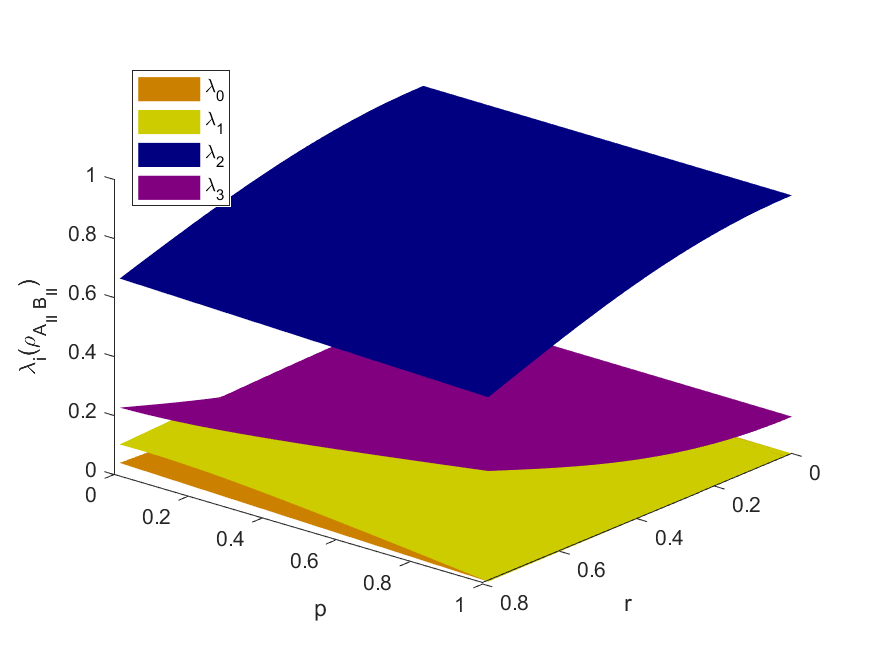}}
\captionsetup{justification=raggedright} 
\caption{Eigenvalues of the reduced states as functions of the state parameter $p$ and Alice's proper acceleration $a_A$, with Bob's proper acceleration fixed at $a_B=2\pi\omega/\ln 3$. The horizontal plotting variable is the dimensionless acceleration parametrization used in the formulas.}
\label{fig:quad1}
\end{figure*}

In the numerical analysis, Alice's proper acceleration $a_A$ is varied, while Bob's proper acceleration is fixed at \(a_B=2\pi\omega/\ln 3\). In the formulas this fixed acceleration corresponds to \(\sin^2 r_B=1/4\) and \(\cos^2 r_B=3/4\). The relative magnitudes of eigenvalues for these four density matrices are illustrated in Fig.\ref{fig:quad1}. 

Their quantum battery capacities Eq. (\ref{batty}) are given by,
\begin{align}
C(\rho_{A_{\text{I}} B_{\text{I}}}, H) &= \sin^2 r + \frac{1}{4} + \sqrt{\left( \sin^2 r - \frac{1}{4} \right)^2 + 3p^2\cos^2 r}, 
\end{align}
\begin{align}
C(\rho_{A_{\text{I}} B_{\text{II}}}, H) &= \sin^2 r + \frac{3}{4} + \sqrt{\left( \sin^2 r - \frac{3}{4} \right)^2 + p^2\cos^2 r}, 
\end{align}
\begin{align}
C(\rho_{A_{\text{II}} B_{\text{I}}}, H) &= \cos^2 r + \frac{1}{4} + \sqrt{\left( \cos^2 r - \frac{1}{4} \right)^2 + 3p^2\sin^2 r}, 
\end{align}
\begin{align}
C(\rho_{A_{\text{II}} B_{\text{II}}}, H) &= \cos^2 r + \frac{3}{4} + \sqrt{\left( \cos^2 r - \frac{3}{4} \right)^2 + p^2\sin^2 r}.
\end{align}

\begin{figure*}[htbp]
    \centering
    \subfigure[]
    {
        \includegraphics[width=5cm]{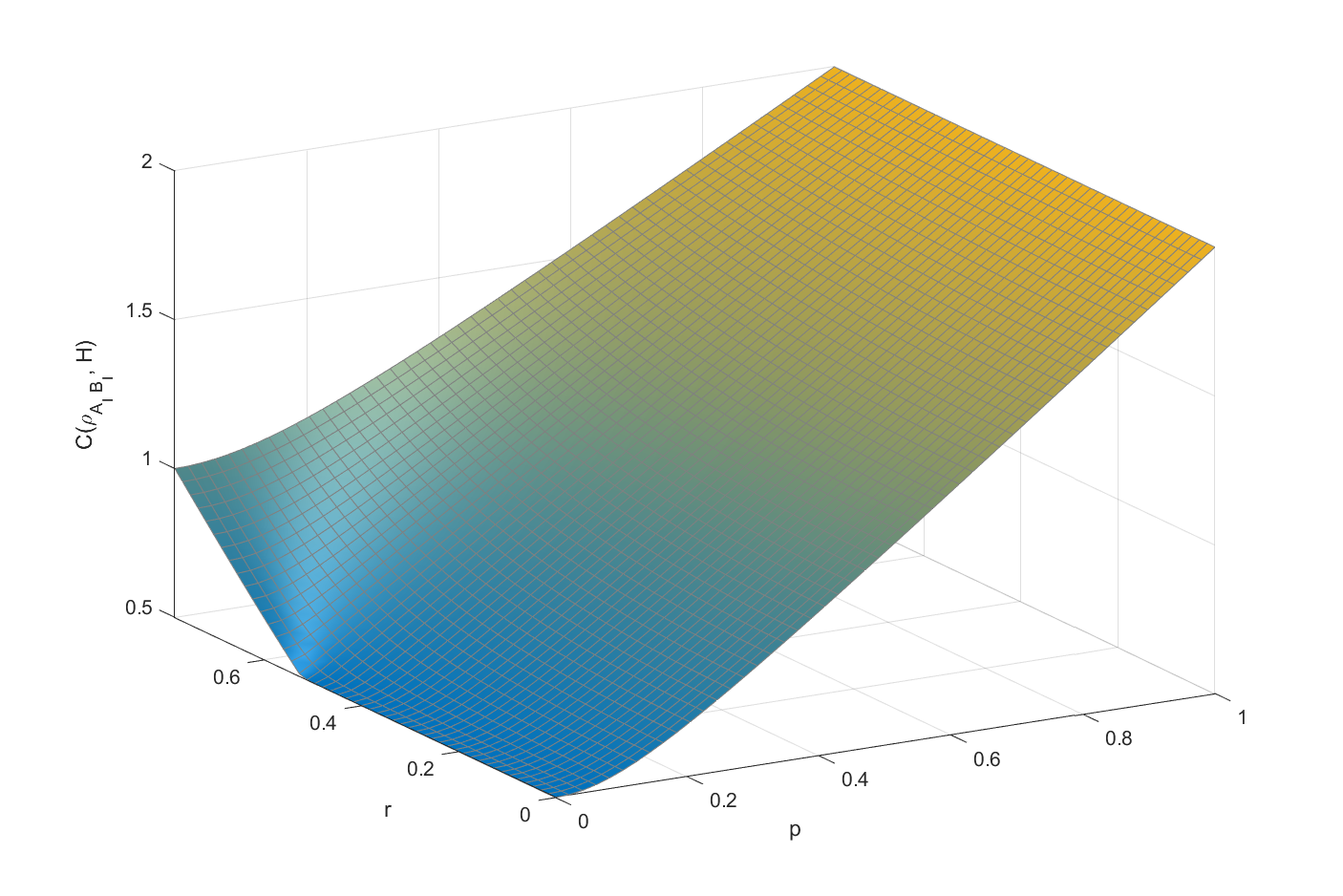}
    }
    \subfigure[]
    {
        \includegraphics[width=5cm]{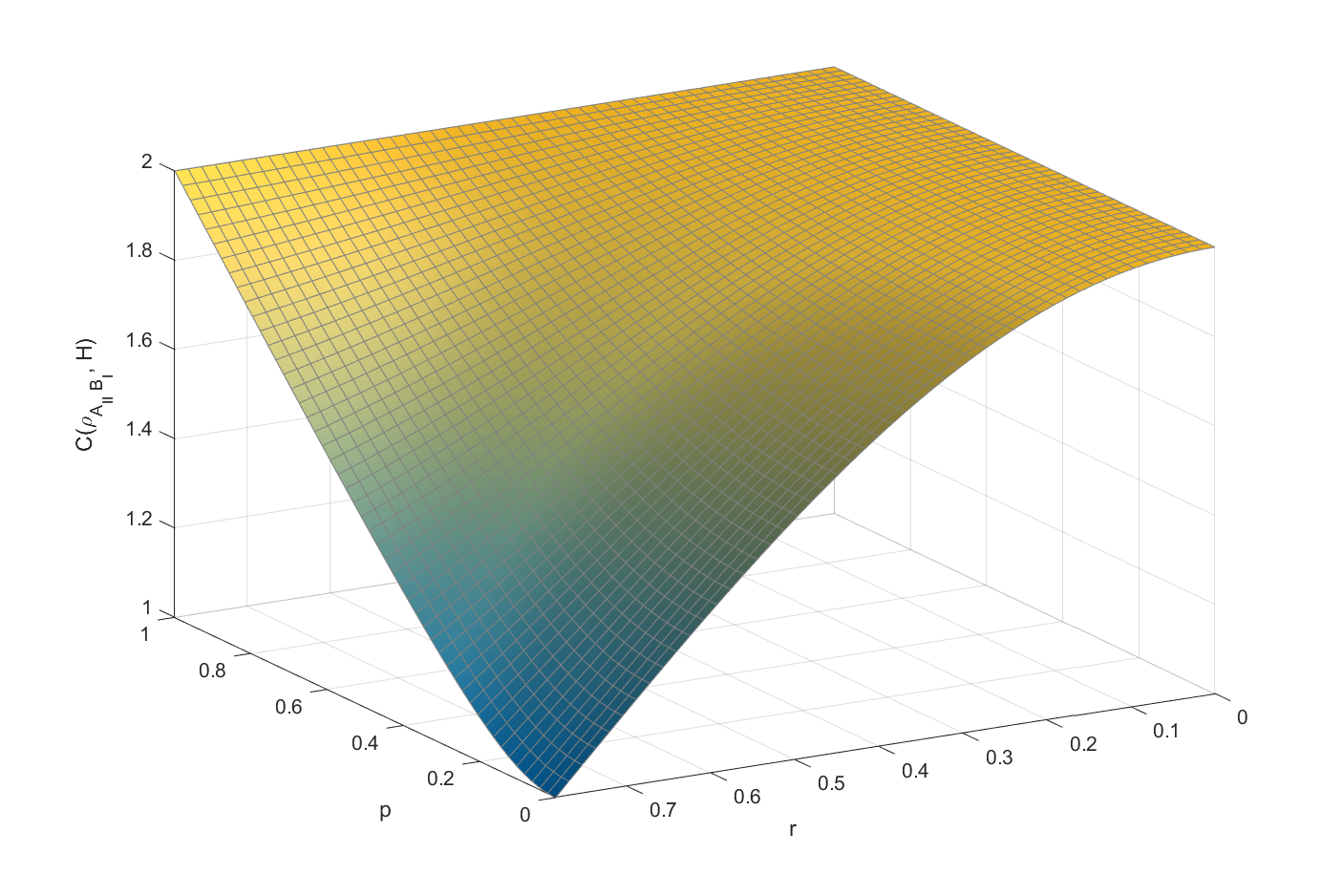}
    }
    \subfigure[]
    {
        \includegraphics[width=5cm]{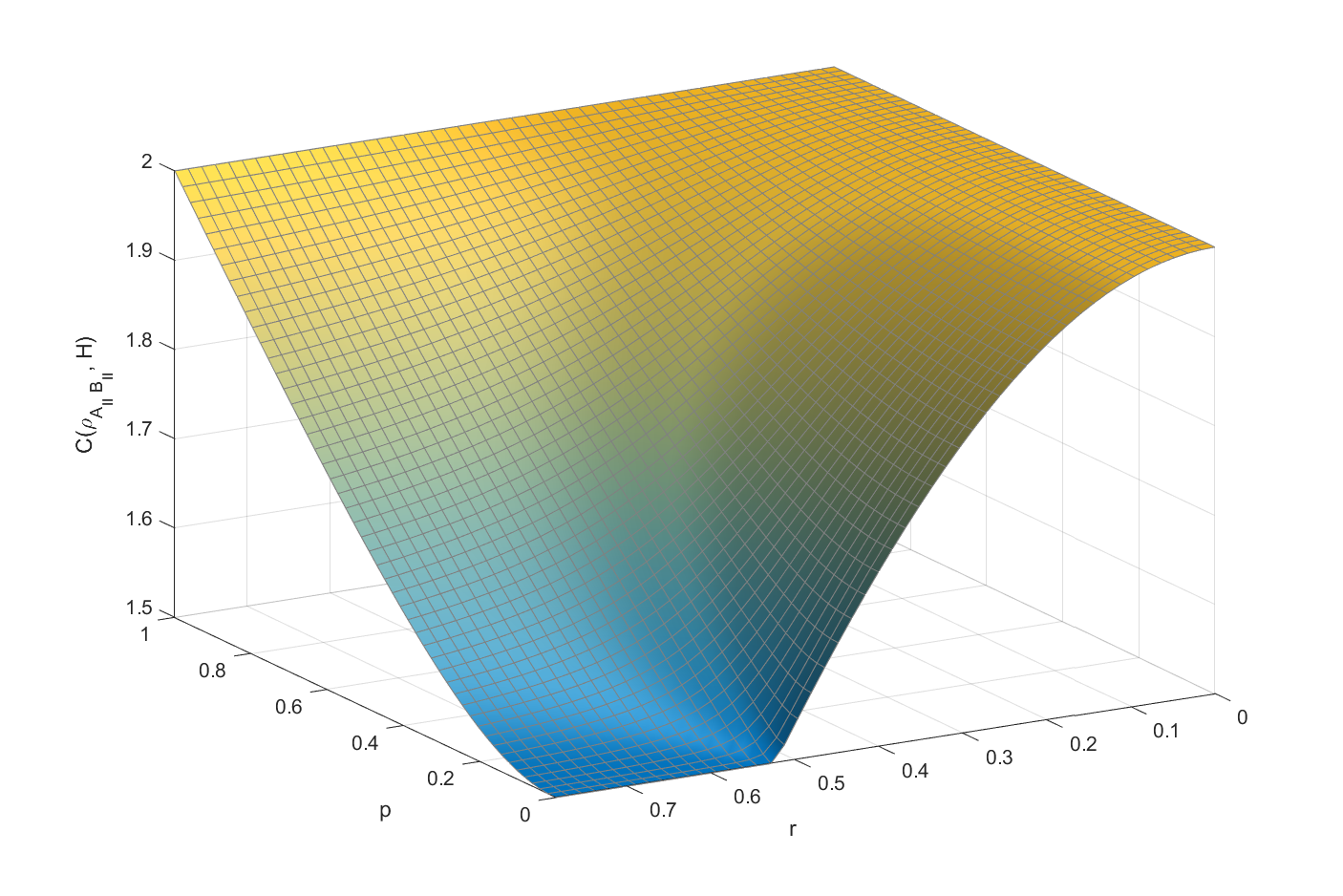}
    } \\
    \subfigure[]
    {
        \includegraphics[width=5cm]{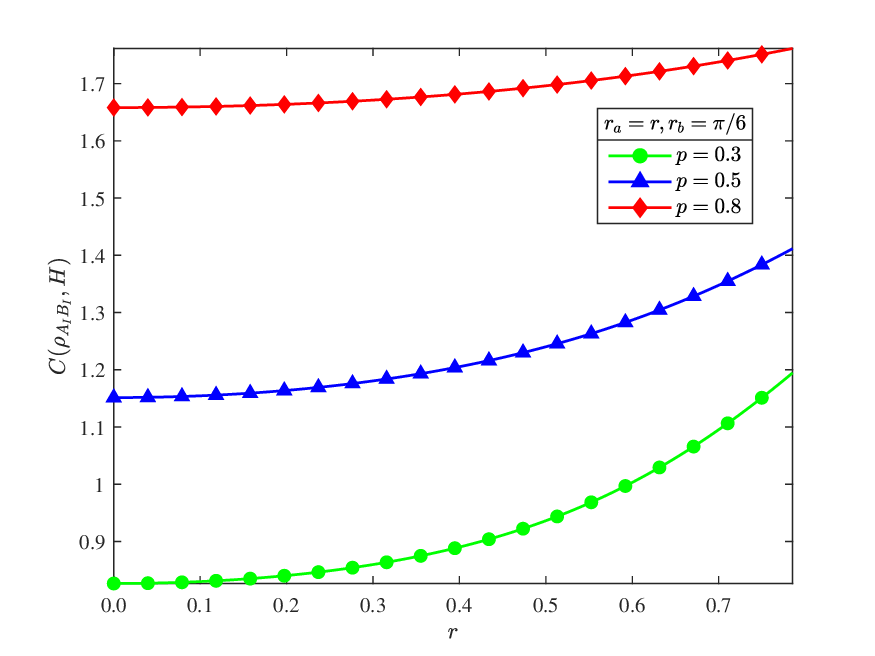}
    }
    \subfigure[]
    {
        \includegraphics[width=5cm]{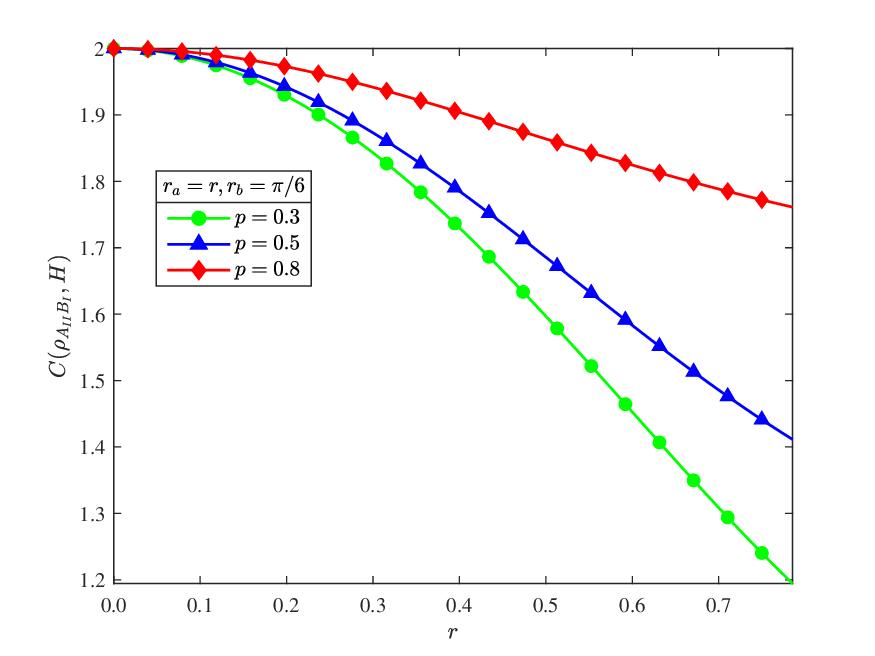}
    }
    \subfigure[]
    {
        \includegraphics[width=5cm]{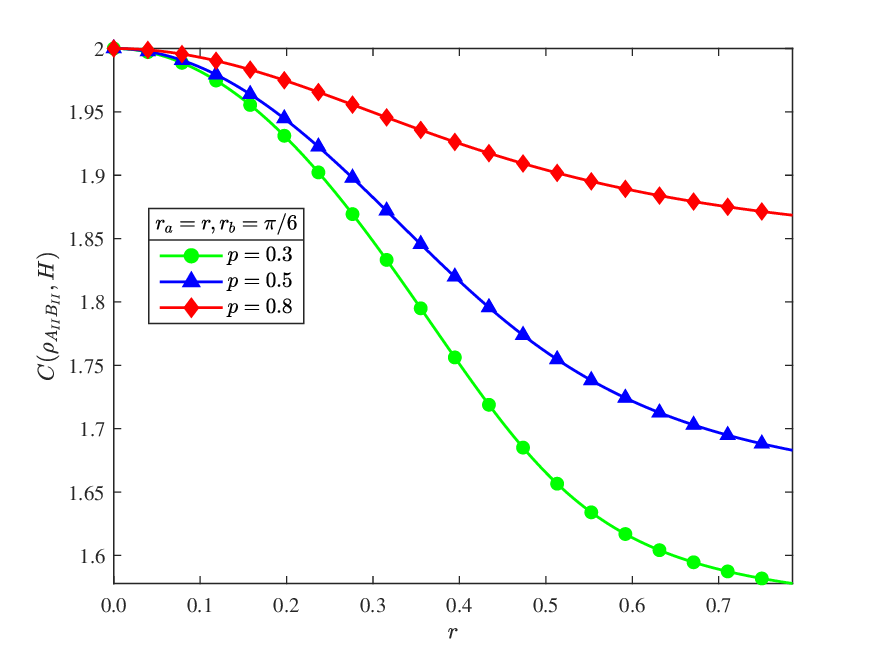}
    }
\captionsetup{justification=raggedright} 
\caption{Quantum battery capacities of the states $\rho_{A_I B_I}$, $\rho_{A_{\text{II}} B_I}$, and $\rho_{A_{\text{II}} B_{\text{II}}}$ as functions of the state parameter $p$ and Alice's proper acceleration $a_A$, with Bob's proper acceleration fixed at $a_B=2\pi\omega/\ln 3$.}
\label{Fig.2}
\end{figure*}

Fig. \ref{Fig.2} shows the quantum battery capacities as functions of Alice's proper acceleration $a_A$ and the state parameter $p$.

As shown in Fig. \ref{Fig.2}(a), when $p=0$ the isotropic state reduces to the maximally mixed state. In this case, the capacity in the accessible sector remains unchanged for accelerations below Bob's fixed acceleration, \(a_A<a_B\), and starts to increase only when \(a_A>a_B\). When $p=1$, the initial state is maximally entangled and the capacity remains maximal over the whole acceleration range. For intermediate values of $p$, the capacity of the physically accessible state, $C(\rho_{A_{\text{I}} B_{\text{I}}},H)$, increases as Alice's acceleration increases. By contrast, the capacities associated with the inaccessible sectors in Figs. \ref{Fig.2}(b) and \ref{Fig.2}(c) generally decrease with increasing acceleration, displaying a trend similar to discharge.

These results indicate that the Unruh channel can enhance the capacity of the battery state accessible in Region I, while the capacity associated with sectors involving inaccessible Region II modes is reduced. This behavior should not be interpreted as a universal protection of quantum resources. Rather, the battery capacity depends on the eigenvalue ordering and population structure of the reduced density matrix relative to the spectrum of $H$. The Unruh channel redistributes the level populations of the accessible state, and for the Ising type Hamiltonian this redistribution can increase the effective population imbalance relevant to the battery capacity.

\section{IV. Quantum Battery Capacity under Noisy Environments}
Quantum states are inherently fragile information carriers. When a quantum state inevitably interacts with environments, such interactions may disrupt its superposition and entanglement properties, a process commonly termed decoherence or quantum noise \cite{ref41, ref42, ref43, ref44, ref45}. 

We next consider the evolution of the battery state when each subsystem is subjected to a local noise channel. After the action of a noisy channel, the state $\rho$ is transformed as
\begin{equation}
\rho^{\prime} =  \sum_i E_i \rho E_i^\dagger, 
\end{equation}
where $E_i$ are the single qubit Kraus operators of the noisy channel and $E_i^\dagger$ denotes the Hermitian conjugate of $E_i$. They satisfy the completeness relation $\sum_{i} E_{i}^{\dagger} E_{i} = I$. Table I summarizes the Kraus operator representations for three archetypal qubit noise channels: phase flip (pf), bit flip (bf), and depolarizing (dep). The corresponding transformations of the Pauli operators, $\sum_i E_i \sigma_k E_i^\dagger$ with $k=1,2,3$, are summarized in Table II. 
\begin{table}
\begin{center}
\begin{tabular}{c c}
\hline \hline
 & $\textrm{Kraus operators}$                                         \\  \hline & \\
pf   & $E_0 = \sqrt{1-k}\, I ,~~~ E_1 = \sqrt{k}\, \sigma_3$                        \\
 & \\
bf   & $E_0 = \sqrt{1-k}\, I ,~~~ E_1 = \sqrt{k} \,\sigma_1$                        \\
 & \\
dep & $E_0 = \sqrt{1-k}\, I ,~~~ E_1 = \sqrt{k/3} \,\sigma_1$                        \\
 & \\
 & $E_2 = \sqrt{k/3}\,\sigma_2 ,~~~ E_3= \sqrt{k/3} \,\sigma_3$                        \\
 & \\
\hline \hline
\end{tabular}
\captionsetup{justification=raggedright} 
\caption[table 1]{Kraus operators for the qubit quantum channels: phase flip channel (pf), bit flip channel (bf), and depolarizing channel (dep), where $k \in [0,1]$ is the decay probability.}
\label{t1}
\end{center}
\end{table}
\begin{table}
    \centering
    \renewcommand{\arraystretch}{1.5}
    \begin{tabular}{cccc}
        \hline \hline
        Channel & $\sigma_1$ & $\sigma_2$ & $\sigma_3$ \\
        \hline
        pf & $(1-2k)\sigma_1$ & $(1-2k)\sigma_2$ & $\sigma_3$ \\
        bf & $\sigma_1$ & $(1-2k)\sigma_2$ & $(1-2k)\sigma_3$ \\
        dep & $(1-4k/3)\sigma_1$ & $(1-4k/3)\sigma_2$ & $(1-4k/3)\sigma_3$ \\
        \hline \hline
    \end{tabular}
    \captionsetup{justification=raggedright}
    \caption{Transformations of Pauli operators under the phase flip channel (pf), bit flip channel (bf), and depolarizing channel (dep).}
    \label{tab:example}
\end{table}

For clarity in the numerical analysis, we assume identical decay probabilities for the two local channels and set $p=0.3$.

\subsection{A. Phase flip channel}
Using Eq. (\ref{batty}), the quantum battery capacities under phase flip noise are
 \begin{figure*}
 \begin{subequations}\label{ph}
 \begin{align}
 C_{pf}(\rho_{\text{A}_{\text{I}}\text{B}_{\text{I}}}, H) 
     &= \sin^2 r + \frac{1}{4} 
     + \sqrt{ \left( \sin^2 r - \frac{1}{4} \right)^2 + \frac{27}{100}(1-2k)^4\cos^2 r },  \\
 C_{pf}(\rho_{\text{A}_{\text{I}}\text{B}_{\text{II}}}, H) 
     &= \sin^2 r + \frac{3}{4} 
     + \sqrt{ \left( \sin^2 r - \frac{3}{4} \right)^2 + \frac{9}{100}(1-2k)^4\cos^2 r },  \\
 C_{pf}(\rho_{\text{A}_{\text{II}}\text{B}_{\text{I}}}, H) 
    &= \cos^2 r + \frac{1}{4} 
    + \sqrt{ \left( \cos^2 r - \frac{1}{4} \right)^2 + \frac{27}{100}(1-2k)^4\sin^2 r }, \\     
 C_{pf}(\rho_{\text{A}_{\text{II}}\text{B}_{\text{II}}}, H) 
     &= \cos^2 r + \frac{3}{4} 
      + \sqrt{ \left( \cos^2 r - \frac{3}{4} \right)^2 + \frac{9}{100}(1-2k)^4\sin^2 r },
 \end{align}
 \end{subequations}
 \noindent \rule[-10pt]{18cm}{0.05em}
 \end{figure*}
where $k \in [0,1]$ is the decay probability for applying a Pauli-$Z$ operation, which models random phase reversals. Eqs. (\ref{ph}) show that the quantum battery capacity is symmetric about $k = \frac{1}{2}$.  

\begin{figure*}[htbp]
\subfigure[]
  {
      \includegraphics[width=5cm]{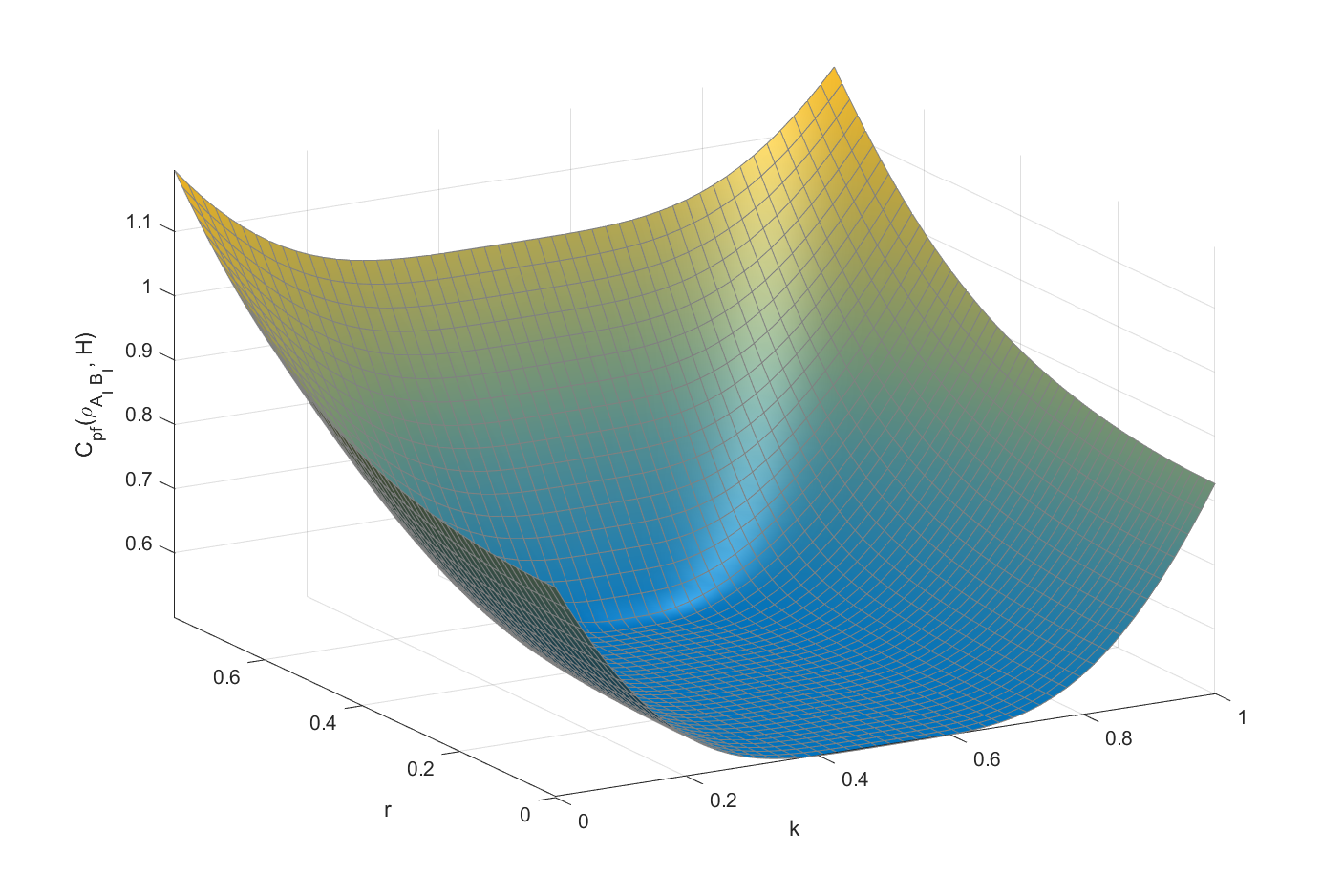}
  }
\subfigure[]
  {
      \includegraphics[width=5cm]{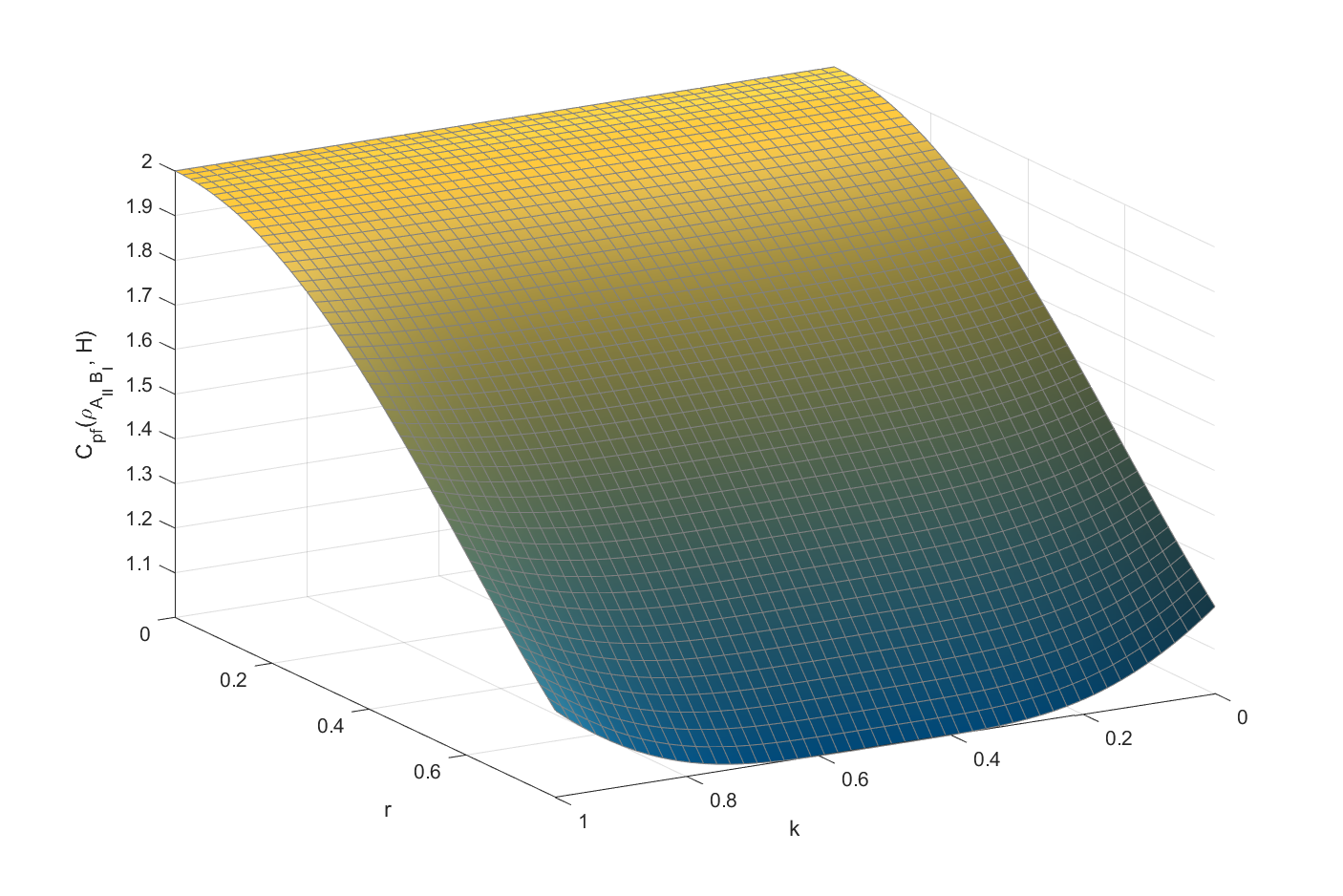}
  }
\subfigure[]
  {
      \includegraphics[width=5cm]{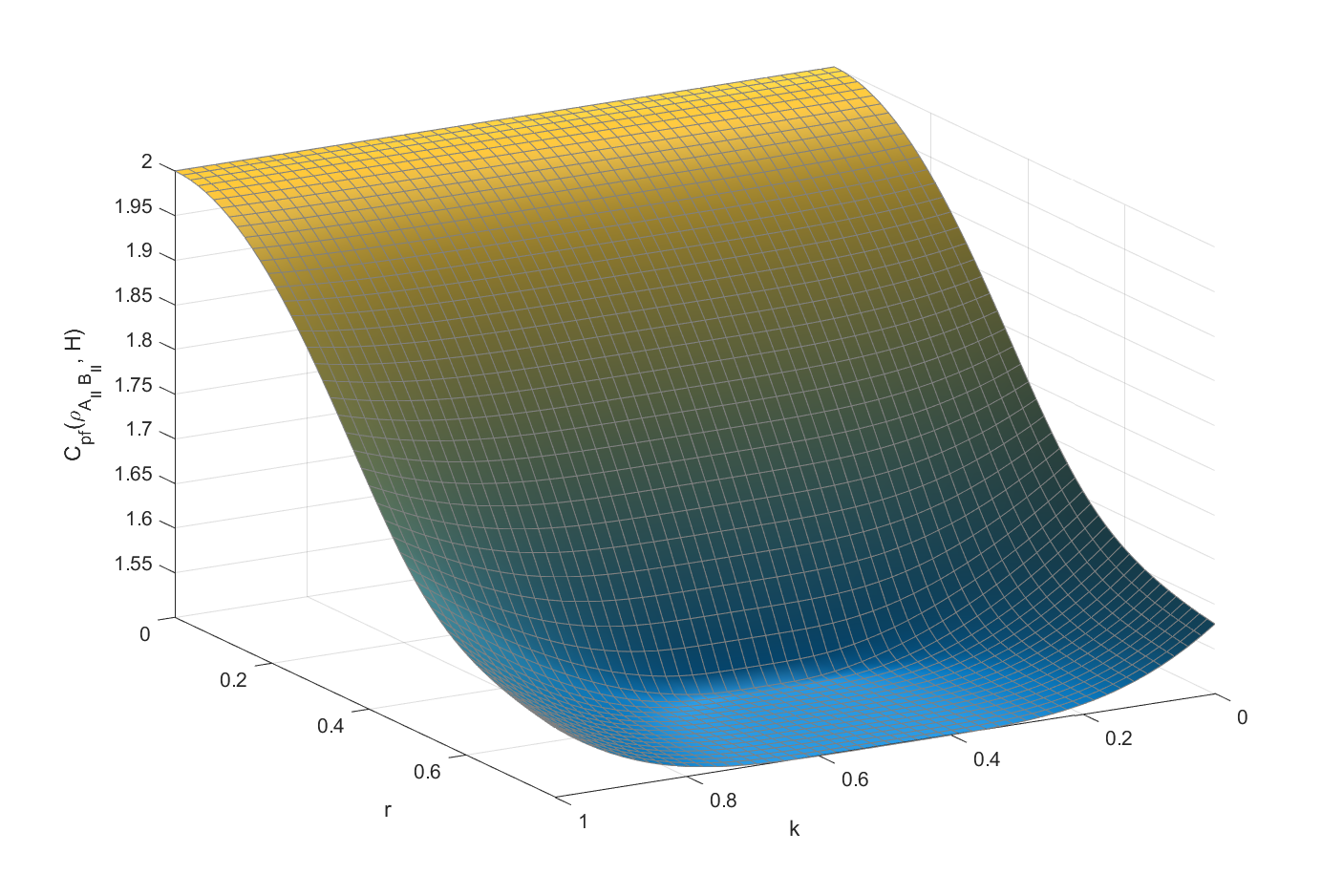}
  }\\
    \subfigure[]
  {
      \includegraphics[width=5cm]{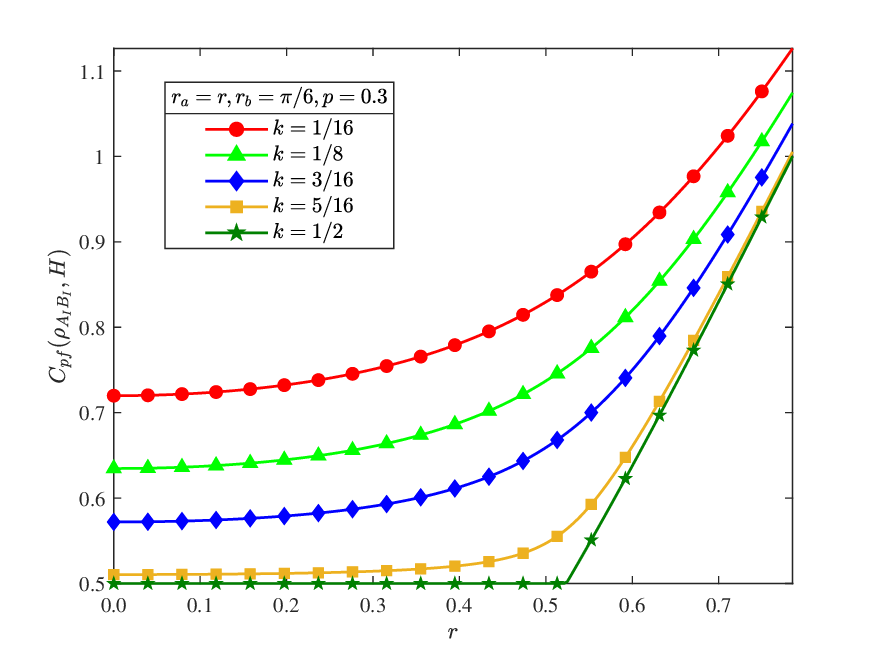}
  }
\subfigure[]
  {
      \includegraphics[width=5cm]{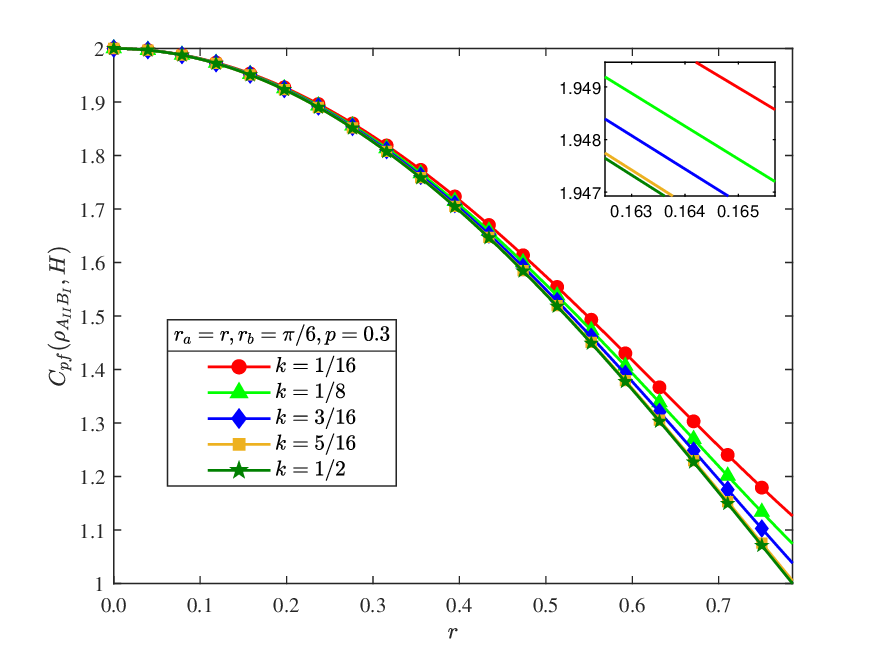}
  }
\subfigure[]
  {
      \includegraphics[width=5cm]{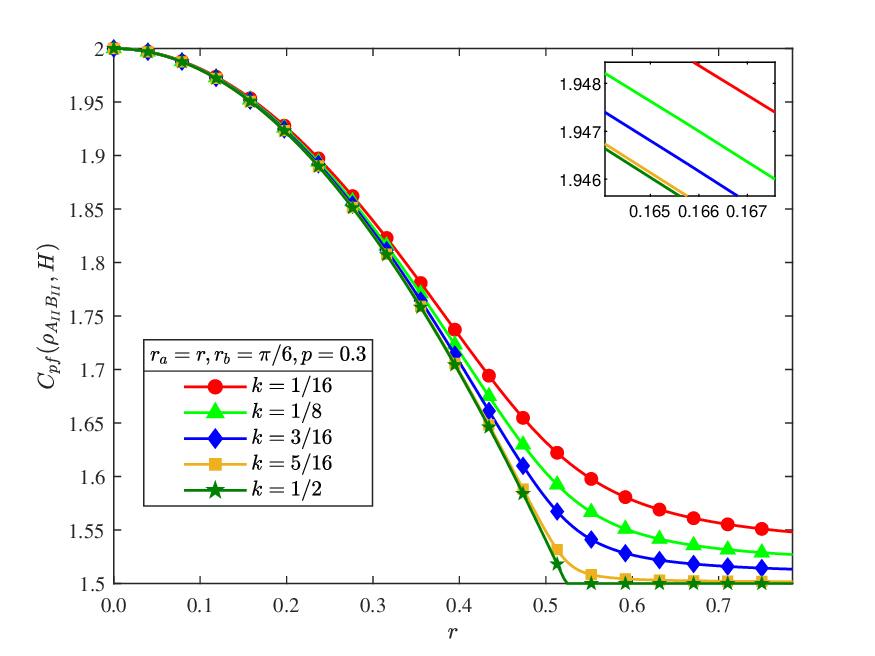}
  }
\captionsetup{justification=raggedright} 
\caption{Quantum battery capacities of the states $\rho^{pf}_{A_I B_I}$, $\rho^{pf}_{A_{\text{II}} B_I}$, and $\rho^{pf}_{A_{\text{II}} B_{\text{II}}}$ under the phase flip channel as functions of the decay probability $k$ and Alice's proper acceleration $a_A$, with Bob's proper acceleration fixed at $a_B=2\pi\omega/\ln 3$ and $p=0.3$.}
\label{fig:quad4}
\end{figure*}

In Fig. \ref{fig:quad4}, we plot the quantum battery capacities as functions of the decay probability $k$ and Alice's proper acceleration $a_A$.

When $k = 0$, the battery capacity reduces to the noise free case. As $k$ increases toward $\frac{1}{2}$ (the point of maximal phase flip noise) the battery capacity decreases monotonically, reaching its minimum at $k = \frac{1}{2}$. The total battery capacities of the physically accessible and inaccessible regions display qualitatively similar dependence on Alice's acceleration: in both regions, the battery capacity is degraded by the presence of noise. The battery capacity of the physically accessible state $C_{pf}(\rho_{\text{A}_{\text{I}}\text{B}_{\text{I}}}, H)$ is strongly suppressed at low acceleration, with the most pronounced decrease occurring for \(a_A<a_B\). For the physically inaccessible state $C_{pf}(\rho_{\text{A}_{\text{II}}\text{B}_{\text{II}}}, H)$, the battery capacity drops rapidly as $a_A$ increases through the region of low acceleration and then levels off for \(a_A>a_B\). Another physically inaccessible state $C_{pf}(\rho_{\text{A}_{\text{II}}\text{B}_{\text{I}}}, H)$ follows a trend similar to the noise free case. Its acceleration dependence is largely unchanged by the phase flip noise (aside from the overall suppression due to $k$). Taken together, both accessible and inaccessible region capacities show qualitatively similar dependence on the Unruh acceleration and are both reduced by increasing phase flip noise, with the strongest sensitivity occurring below Bob's fixed acceleration.

\subsection{B. Bit flip channel}
The quantum battery capacities under the bit flip channel are given by Eqs. (\ref{bf}).

\begin{figure*}
 \begin{subequations}\label{bf}
\begin{align}
C_{bf}(\rho_{\text{A}_{\text{I}}\text{B}_{\text{I}}}, H)=&
    \sqrt{(1 - 2k)^2\left( \sin^2 r + \frac{1}{4} \right)^2 + \frac{27}{400}\left(1-(1 - 2k)^2\right)^2\cos^2 r} \\& +  
    \sqrt{(1 - 2k)^2\left( \sin^2 r - \frac{1}{4} \right)^2 + \frac{27}{400}\left(1+(1 - 2k)^2\right)^2\cos^2 r},   \notag \\
C_{bf}(\rho_{\text{A}_{\text{I}}\text{B}_{\text{II}}}, H)=& 
    \sqrt{(1 - 2k)^2\left( \sin^2 r + \frac{3}{4} \right)^2 + \frac{9}{400}\left(1-(1 - 2k)^2\right)^2\cos^2 r}\\&+  
     \sqrt{(1 - 2k)^2\left( \sin^2 r - \frac{3}{4} \right)^2 + \frac{9}{400}\left(1+(1 - 2k)^2\right)^2\cos^2 r}, \notag \\
C_{bf}(\rho_{\text{A}_{\text{II}}\text{B}_{\text{I}}}, H)=& 
     \sqrt{(1 - 2k)^2\left( \cos^2 r + \frac{1}{4} \right)^2 + \frac{27}{400}\left(1-(1 - 2k)^2\right)^2\sin^2 r}\\&+  
    \sqrt{(1 - 2k)^2\left( \cos^2 r - \frac{1}{4} \right)^2 + \frac{27}{400}\left(1+(1 - 2k)^2\right)^2\sin^2 r},   \notag \\
C_{bf}(\rho_{\text{A}_{\text{II}}\text{B}_{\text{II}}}, H)=&
    \sqrt{(1 - 2k)^2\left( \cos^2 r + \frac{3}{4} \right)^2 + \frac{9}{400}\left(1-(1 - 2k)^2\right)^2\sin^2 r}\\&+  
    \sqrt{(1 - 2k)^2\left( \cos^2 r - \frac{3}{4} \right)^2 + \frac{9}{400}\left(1+(1 - 2k)^2\right)^2\sin^2 r}.     \notag 
\end{align}
\end{subequations}
\noindent \rule[-10pt]{18cm}{0.05em}
\end{figure*}



\begin{figure*}[htbp]
\subfigure[]
  {
      \includegraphics[width=5cm]{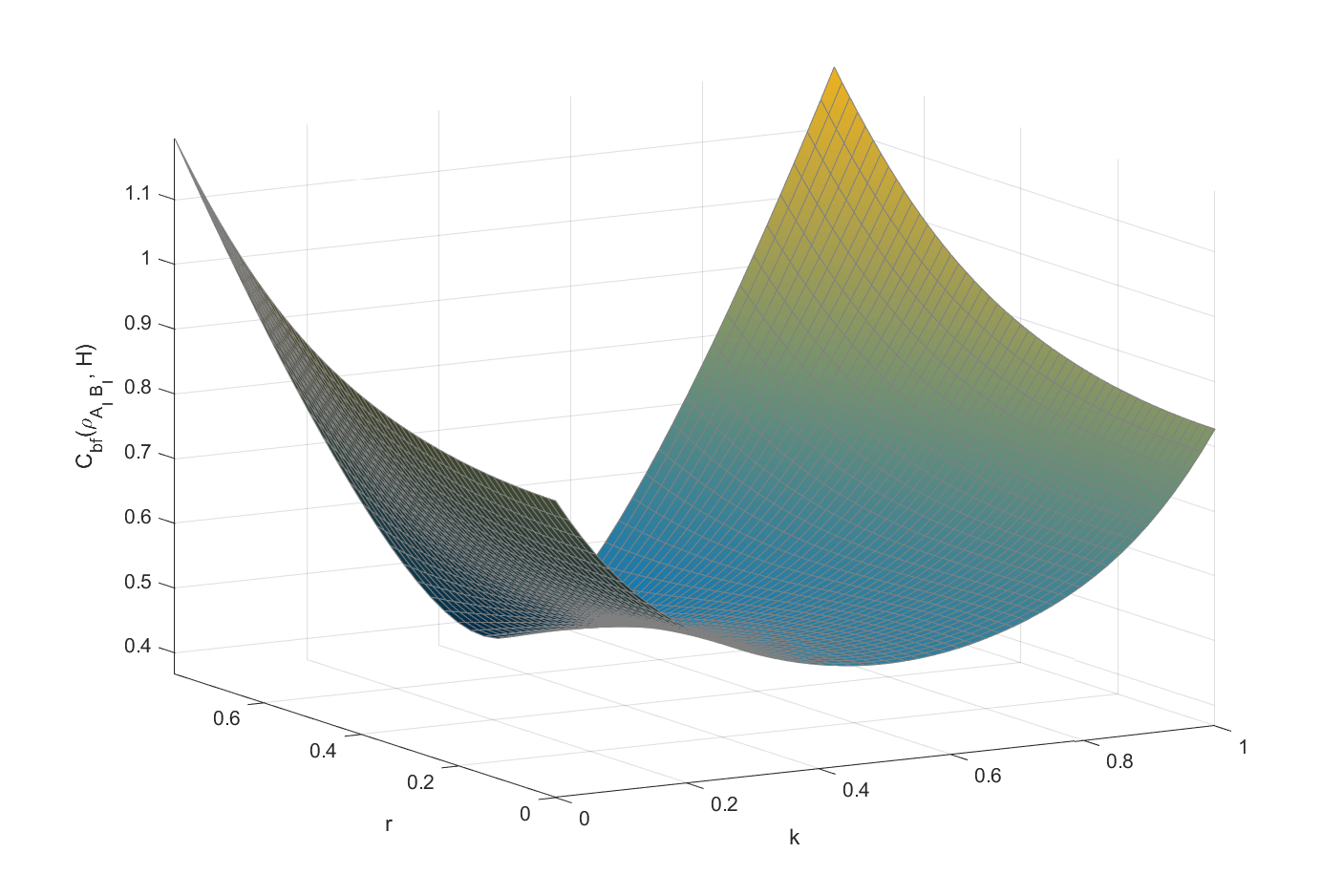}
  }
\subfigure[]
  {
      \includegraphics[width=5cm]{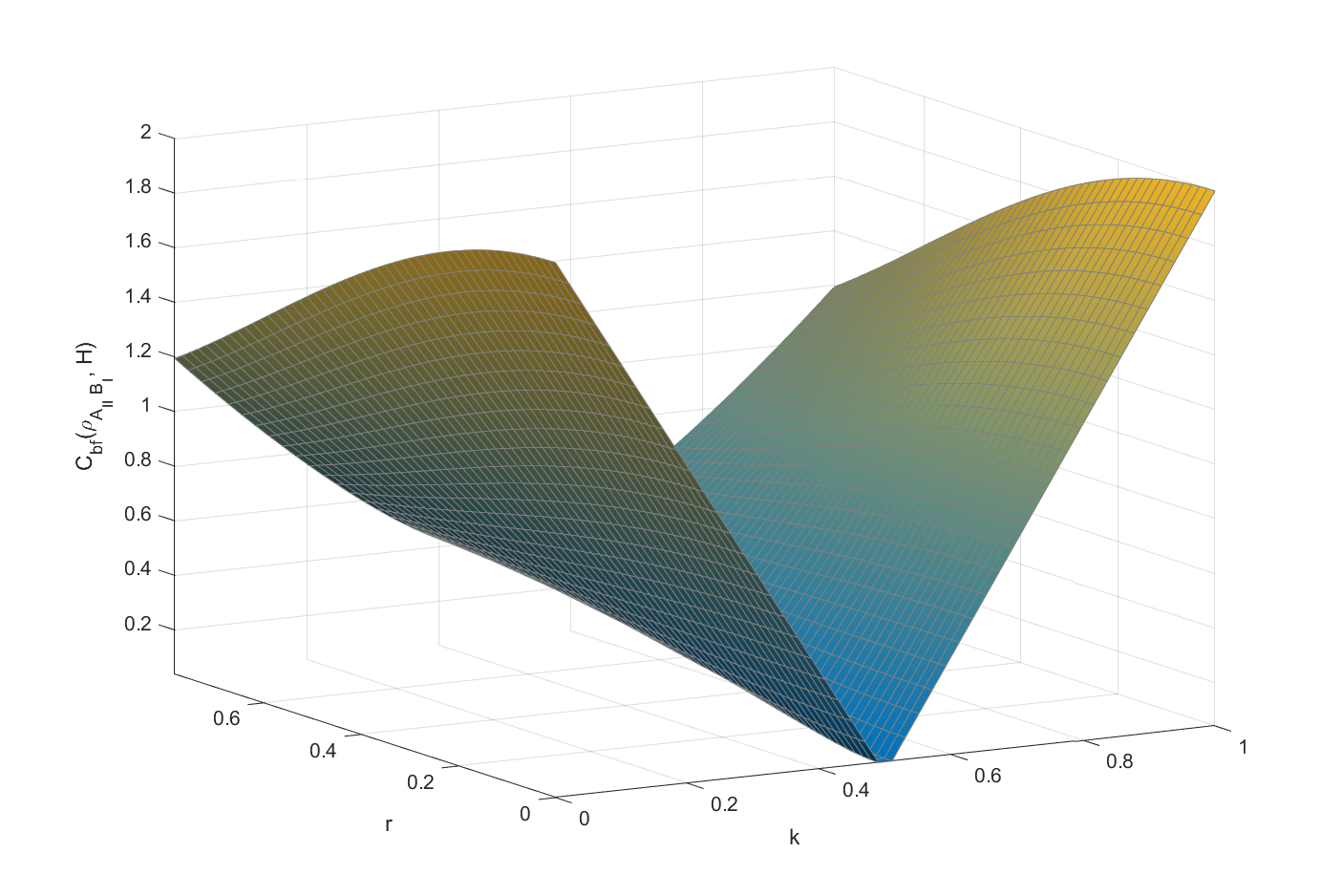}
  }
\subfigure[]
  {
      \includegraphics[width=5cm]{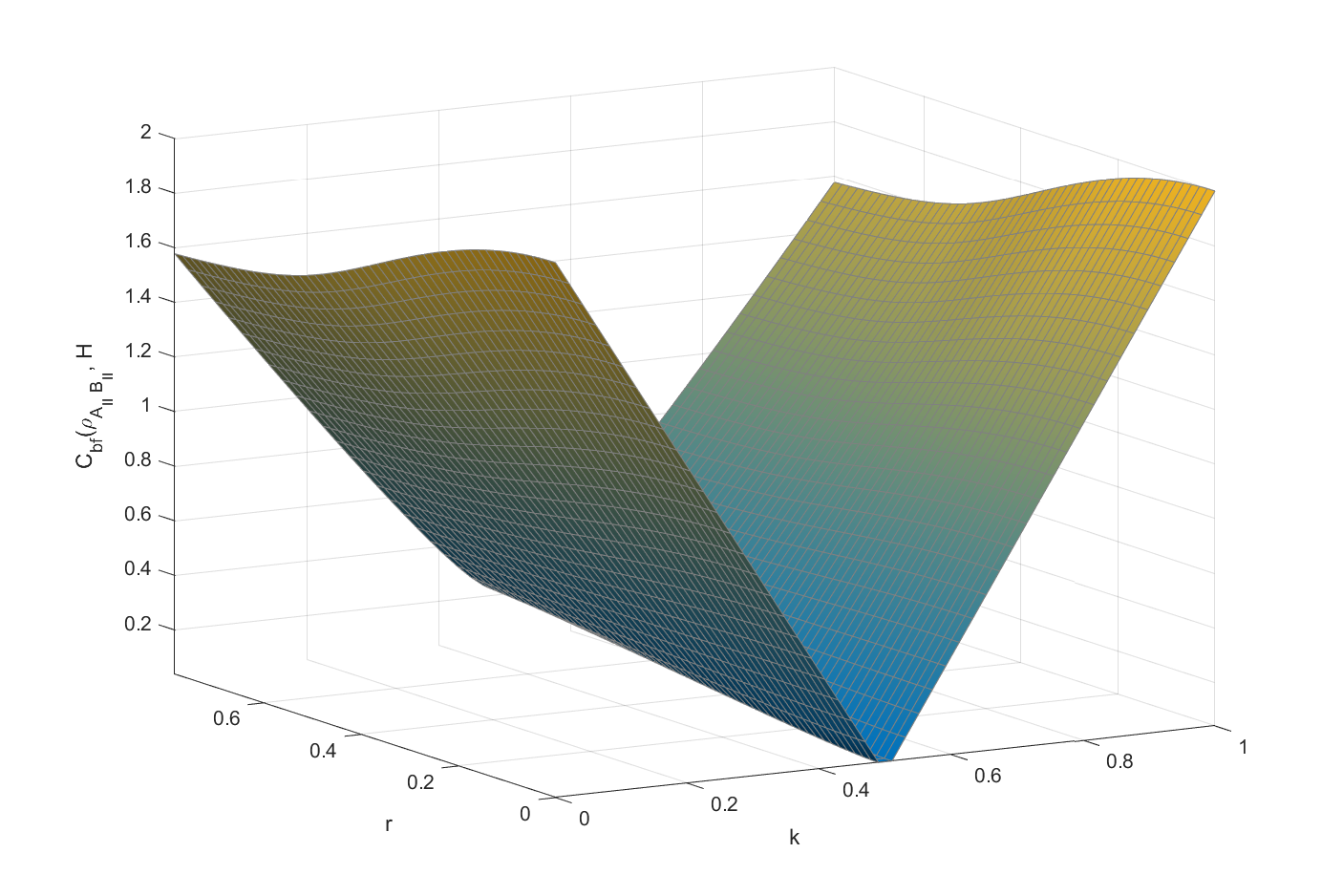}
  }\\
    \subfigure[]
  {
      \includegraphics[width=5cm]{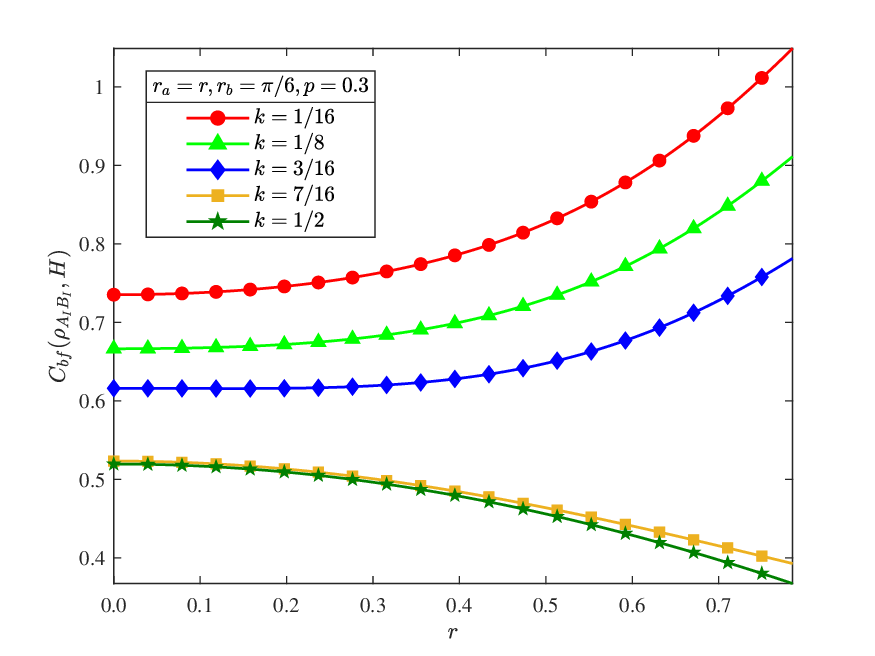}
  }
\subfigure[]
  {
      \includegraphics[width=5cm]{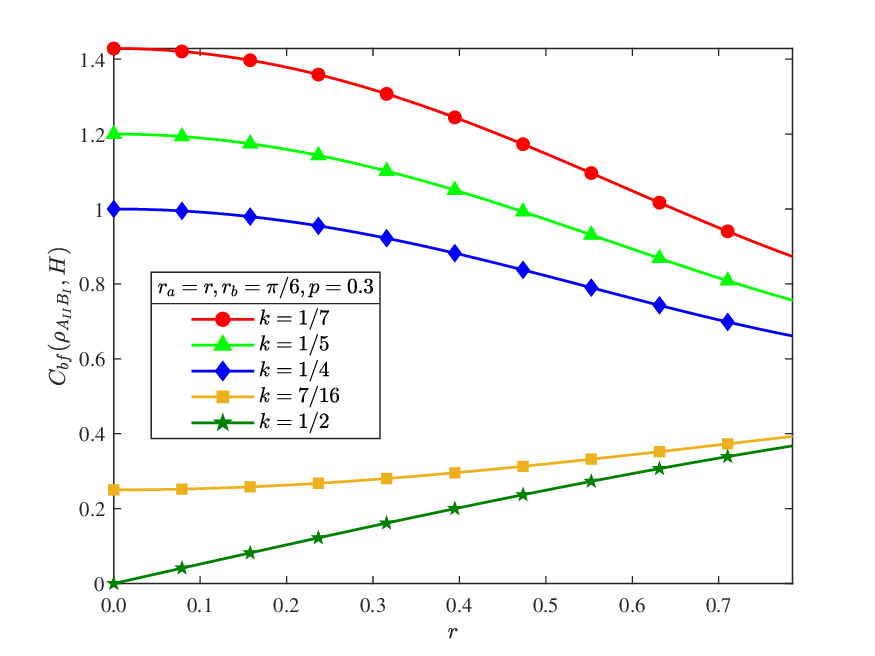}
  }
\subfigure[]
  {
      \includegraphics[width=5cm]{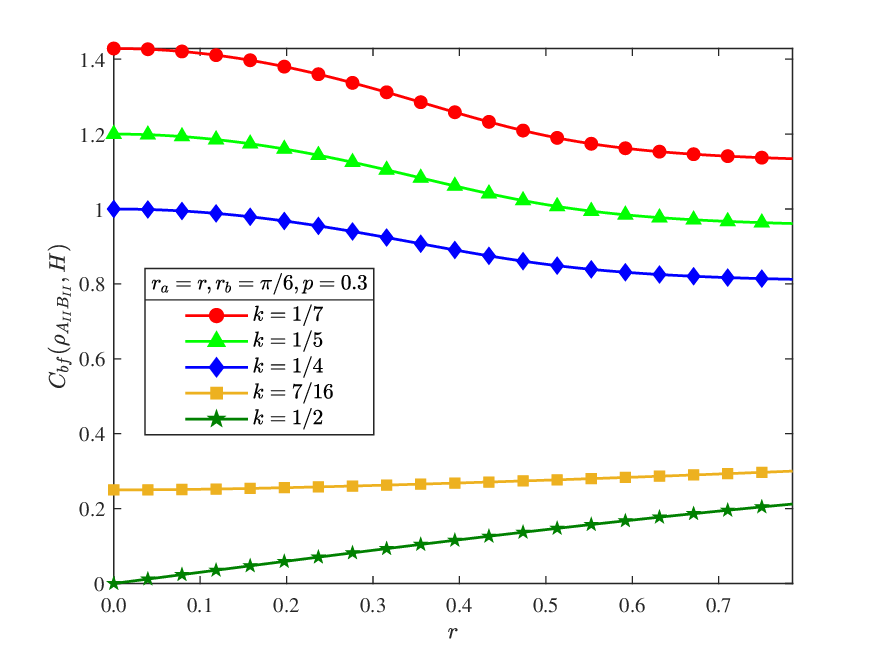}
  }
\captionsetup{justification=raggedright} 
\caption{Quantum battery capacities of the states $\rho^{bf}_{A_I B_I}$, $\rho^{bf}_{A_{\text{II}} B_I}$, and $\rho^{bf}_{A_{\text{II}} B_{\text{II}}}$ under the bit flip channel as functions of the decay probability $k$ and Alice's proper acceleration $a_A$, with Bob's proper acceleration fixed at $a_B=2\pi\omega/\ln 3$ and $p=0.3$.}
\label{fig:quad5}
\end{figure*}

We model bit flip noise by a bit flip channel with decay probability $k \in [0,1]$: with probability $k$, a Pauli $X$ operator is applied, producing random bit flips. The resulting quantum battery capacity depends on both Alice's proper acceleration $a_A$ and the noise strength $k$, as shown in Fig. \ref{fig:quad5}. 

Eqs. (\ref{bf}) show that the capacity is symmetric about $k = \frac{1}{2}$. When $k=0$, the result reduces to the noiseless case. As $k$ approaches $\frac{1}{2}$, which is the maximally random bit flip point, the dependence of the capacity on Alice's acceleration becomes less monotonic. The charging behavior of $C_{bf}(\rho_{\text{A}_{\text{I}}\text{B}_{\text{I}}}, H)$ and the behavior similar to discharge of $C_{bf}(\rho_{\text{A}_{\text{II}}\text{B}_{\text{I}}}, H)$ and $C_{bf}(\rho_{\text{A}_{\text{II}}\text{B}_{\text{II}}}, H)$ are correspondingly weakened. This indicates that bit flip noise changes the population distribution of the energy levels more directly than phase flip noise, thereby modifying the charging and discharging trends produced by the Unruh channel.

\subsection{C. Depolarizing channel}
Under depolarizing noise, the quantum battery capacities are given by Eqs. (\ref{dep}).
\begin{figure*}
\begin{subequations}\label{dep}
\begin{align}
C_{dep}(\rho_{\mathrm{A}_{\mathrm{I}}\mathrm{B}_{\mathrm{I}}},H) &
=\Big\lvert 1-\frac{4k}{3}\Big\rvert\Bigg[ \sin^2 r + \frac{1}{4}+ \sqrt{\left(\sin^2 r-\frac{1}{4}\right)^{2}
    + \frac{27}{100}\Big(1-\frac{4k}{3}\Big)^{2}\cos^2 r} \Bigg],\\
C_{dep}\big(\rho_{\mathrm{A}_{\mathrm{I}}\mathrm{B}_{\mathrm{II}}},H\big) &= \Big\lvert 1-\frac{4k}{3}\Big\rvert\Bigg[ \sin^2 r + \frac{3}{4} 
     + \sqrt{\left(\sin^2 r-\frac{3}{4}\right)^{2}
    + \frac{9}{100}\Big(1-\frac{4k}{3}\Big)^{2}\cos^2 r}
    \Bigg],\\
C_{dep}\big(\rho_{\mathrm{A}_{\mathrm{II}}\mathrm{B}_{\mathrm{I}}},H\big)
    &= \Big\lvert 1-\frac{4k}{3}\Big\rvert\Bigg[ \cos^{2} r + \frac{1}{4} 
    + \sqrt{\left(\cos^{2} r-\frac{1}{4}\right)^{2}
    + \frac{27}{100}\Big(1-\frac{4k}{3}\Big)^{2}\sin^{2} r}
    \Bigg],   \\
 C_{dep}\big(\rho_{\mathrm{A}_{\mathrm{II}}\mathrm{B}_{\mathrm{II}}},H\big)
   & = \Big\lvert 1-\frac{4k}{3}\Big\rvert\Bigg[ \cos^{2} r + \frac{3}{4} 
    + \sqrt{\left(\cos^{2} r-\frac{3}{4}\right)^{2}
    + \frac{27}{100}\Big(1-\frac{4k}{3}\Big)^{2}\sin^{2} r}
    \Bigg].
\end{align}
\end{subequations}
\noindent \rule[-10pt]{18cm}{0.05em}
\end{figure*}





\begin{figure*}[htbp]
\subfigure[]
  {
      \includegraphics[width=5cm]{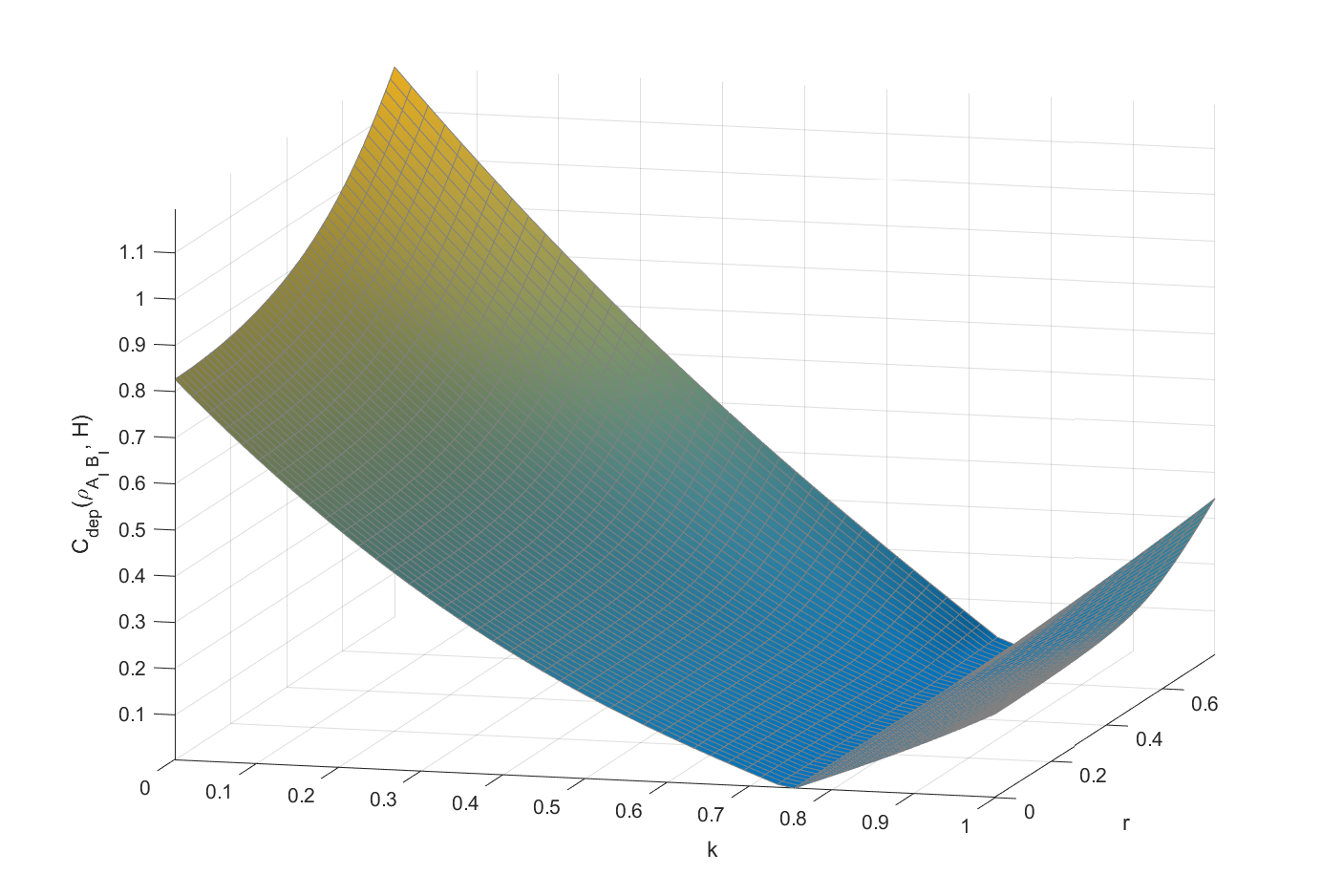}
  }
\subfigure[]
  {
      \includegraphics[width=5cm]{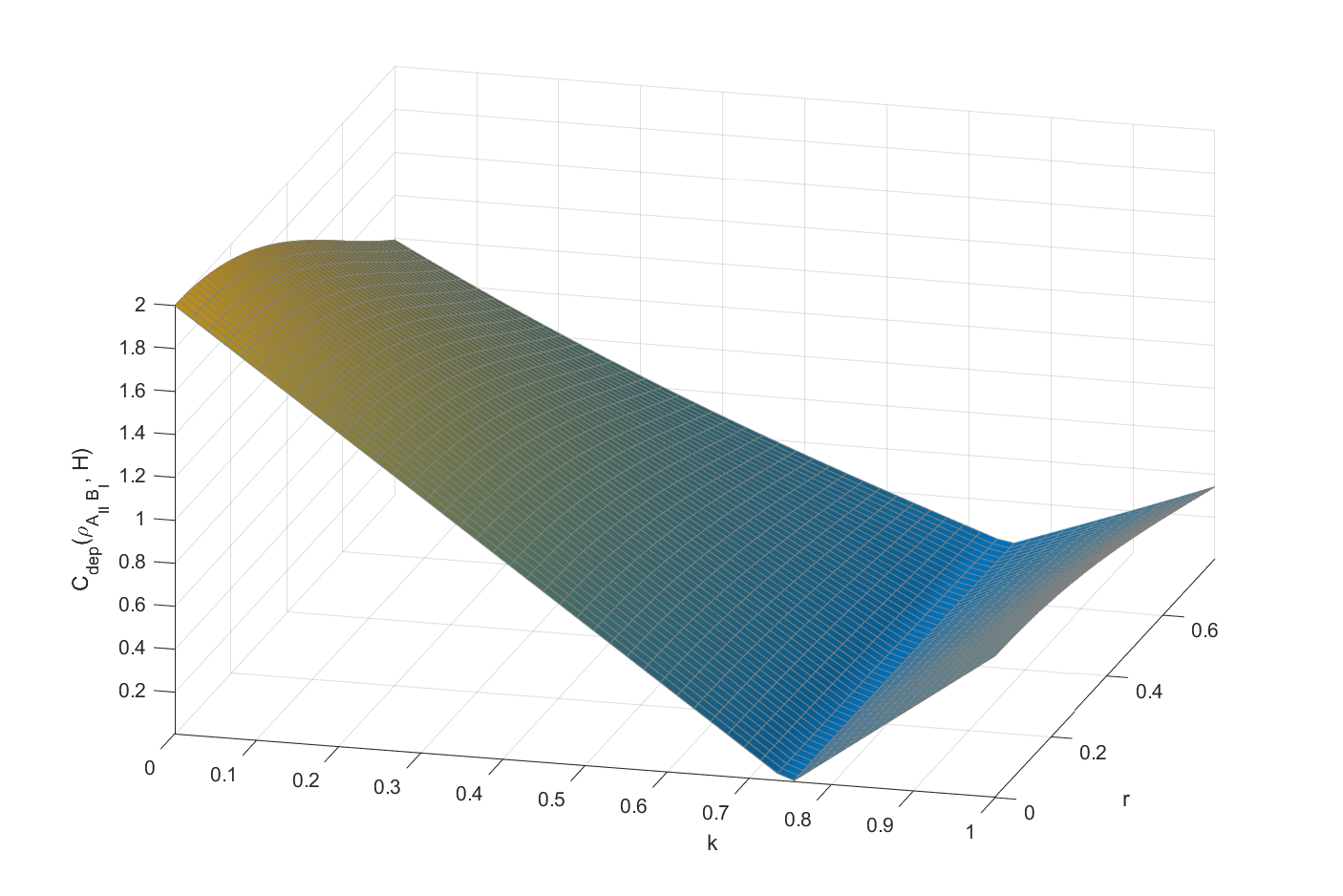}
  }
\subfigure[]
  {
      \includegraphics[width=5cm]{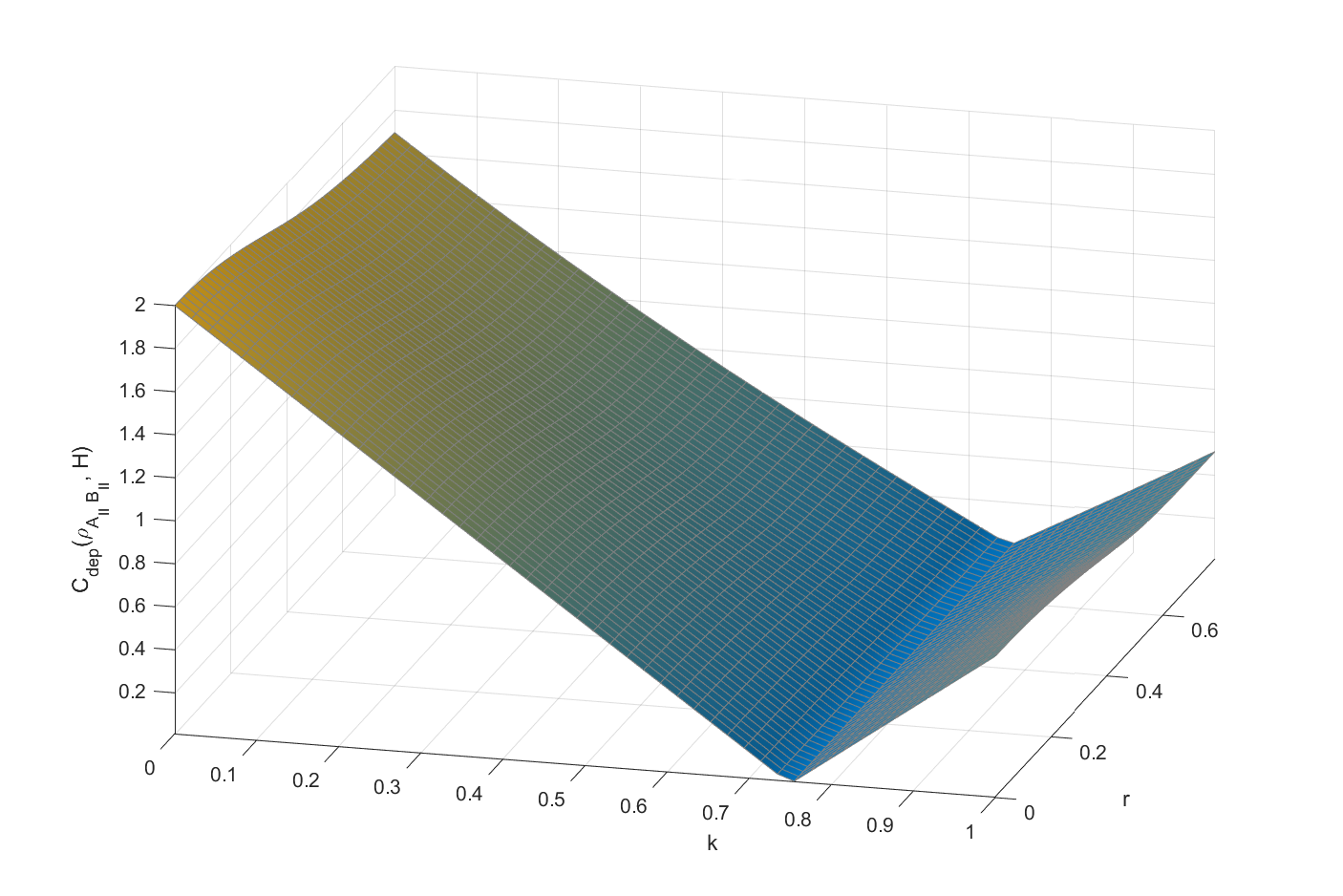}
  }\\
    \subfigure[]
  {
      \includegraphics[width=5cm]{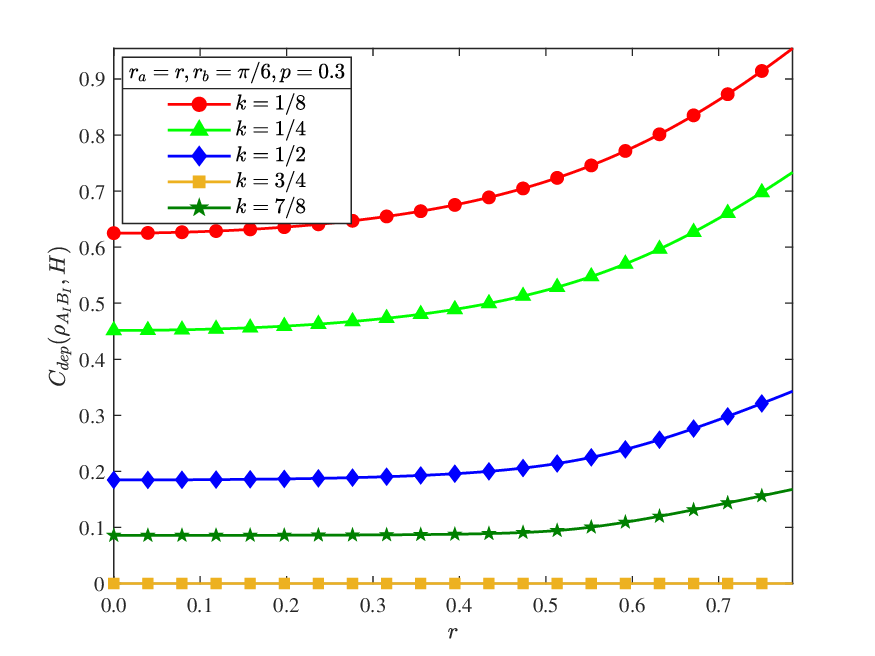}
  }
\subfigure[]
  {
      \includegraphics[width=5cm]{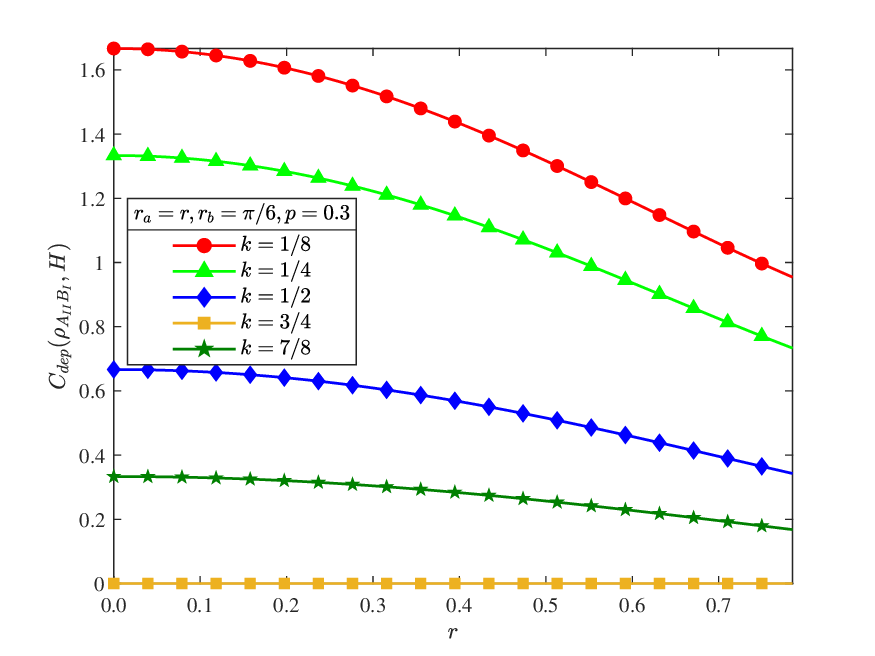}
  }
\subfigure[]
  {
      \includegraphics[width=5cm]{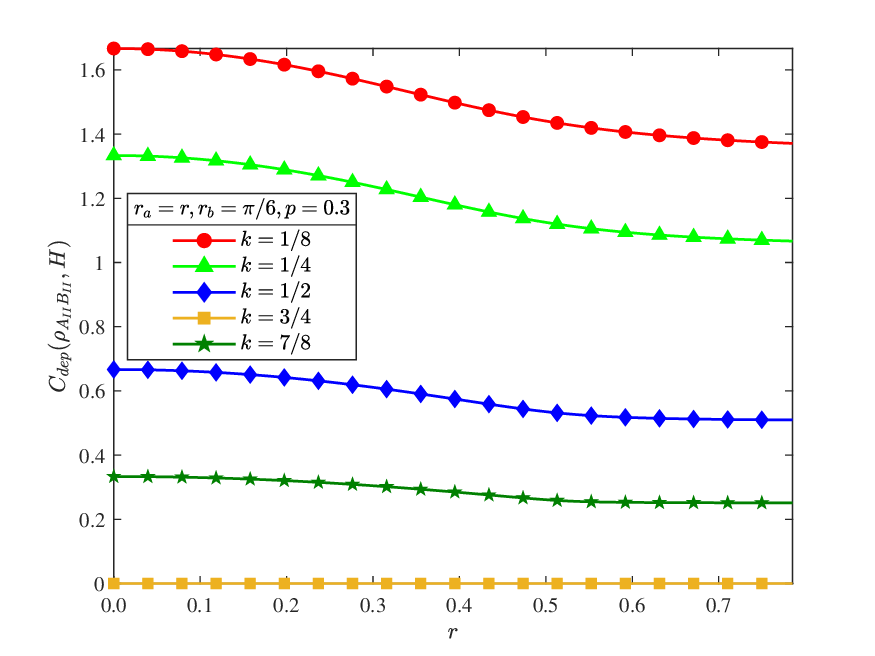}
  }
\captionsetup{justification=raggedright} 
\caption{Quantum battery capacities of the states $\rho^{dep}_{A_I B_I}$, $\rho^{dep}_{A_{\text{II}} B_I}$, and $\rho^{dep}_{A_{\text{II}} B_{\text{II}}}$ under the depolarizing channel as functions of the decay probability $k$ and Alice's proper acceleration $a_A$, with Bob's proper acceleration fixed at $a_B=2\pi\omega/\ln 3$ and $p=0.3$.}
\label{fig:quad6}
\end{figure*}

To account for depolarizing noise, we denote by $k \in [0,1]$ the decay probability such that a single qubit state is transformed to $I/2$ with probability $k$; equivalently, a Pauli $X$, $Y$ or $Z$ operation is applied with probability $k/3$. The parameter $k$ therefore quantifies the degree of depolarization, which attenuates both coherence and polarization. The battery capacity consequently depends on both Alice's proper acceleration $a_A$ and the noise strength $k$, as shown in Fig. \ref{fig:quad6}.

When $k = 0$, the quantum battery capacity reduces to the noiseless case. As $k$ increases toward $\frac{3}{4}$, the monotonic dependence of the capacity on Alice's acceleration remains qualitatively similar to that in the noiseless case, but its amplitude is suppressed by the noise, and the capacity exhibits abrupt death at $k = \frac{3}{4}$. For further increases of $k$ toward $1$, the charging behavior of $C_{dep}(\rho_{\text{A}_{\text{I}}\text{B}_{\text{I}}}, H)$ again mirrors the discharging characteristics of $C_{dep}(\rho_{\text{A}_{\text{II}}\text{B}_{\text{I}}}, H)$ and $C_{dep}(\rho_{\text{A}_{\text{II}}\text{B}_{\text{II}}}, H)$. Overall, the capacities in the physically accessible and inaccessible regions exhibit qualitatively similar acceleration dependence, and they converge to the noiseless behavior except that they vanish abruptly at the maximal noise $k = \frac{3}{4}$.

\section{V. Conclusions and discussions}
We have investigated the influence of the Unruh channel on the capacity of a two cell spin quantum battery in a noninertial relativistic setting. The battery state is modeled by a two qubit isotropic state, and its energetic structure is specified by the Ising type Hamiltonian $H=\sigma_3\otimes\sigma_3$. Within the channel framework, the Rindler modes accessible in Region I constitute the effective battery subsystem, while the causally inaccessible Region II modes play the role of an environment associated with acceleration.

Our results show that increasing the proper acceleration can enhance the capacity of the accessible battery state. This enhancement does not imply that acceleration protects all quantum resources. Instead, it arises because the capacity considered here is determined by the eigenvalue ordering and population structure of the reduced battery state. The Unruh channel and the trace over inaccessible modes can redistribute the populations among the energy levels; for the pure Ising Hamiltonian adopted in this work, this redistribution may increase the effective energy level imbalance and hence enhance the battery capacity.

For noisy environments, we have investigated the evolution of quantum battery capacity under phase flip, bit flip and depolarizing channels. The three noise channels have distinct effects. Phase flip noise mainly suppresses contributions related to coherence, while bit flip noise directly changes the population structure and can therefore alter the charging and discharging trends more strongly. Under depolarizing noise, the capacity is globally suppressed and vanishes at the fully depolarizing point.

A complementary localized detector field model along accelerated trajectories, formulated through an explicit interaction Hamiltonian between battery degrees of freedom and quantum fields, would provide a more dynamical description and is left for future work.

Overall, this study elucidates the interplay between the Unruh channel and environmental noise on quantum battery capacity, providing new insights into the charging and discharging behaviour of quantum batteries in noninertial and noisy settings.

{\bf Acknowledgements:} ~ This work is supported by the National Natural Science Foundation of China (NSFC) under Grant Nos 12564048 and 12371132; the Fundamental Research Funds for the Central Universities; the
China Scholarship Council (CSC); the Natural Science Foundation of Hainan Province under Grant No. 125RC744 and the specific research fund of the Innovation Platform for Academicians of Hainan Province.

\appendix
\onecolumngrid

\section{Appendix A}

Without an additional noise channel, the Bloch representation and eigenvalues of quantum states $\rho_{A_{\text{I}} B_{\text{I}}}$, $\rho_{A_{\text{I}} B_{\text{II}}}$, $\rho_{A_{\text{II}} B_{\text{I}}}$ and $\rho_{A_{\text{II}} B_{\text{II}}}$ have the following expressions, respectively. 

The Bloch representation of quantum state $\rho_{A_{\text{I}} B_{\text{I}}}$: 
\begin{align*}
\rho_{A_{\text{I}} B_{\text{I}}} & = \frac{1}{4} \left( I \otimes I - \sin^2 r_A \sigma_3 \otimes I - \sin^2 r_B I \otimes \sigma_3 + p \cos r_A \cos r_B \sigma_1 \otimes \sigma_1 \right. \nonumber \\
& \quad \left. + p \cos r_A \cos r_B \sigma_2 \otimes \sigma_2 + (\sin^2 r_A \sin^2 r_B - p \cos^2 r_A \cos^2 r_B) \sigma_3 \otimes \sigma_3 \right), 
\end{align*}
the eigenvalues are as follows:
\begin{align*}
\lambda_0 &= \frac{1}{4} \left(1 - \sin^2 r_A - \sin^2 r_B + \sin^2 r_A \sin^2 r_B - p \cos^2 r_A \cos^2 r_B\right), \\
\lambda_1 &= \frac{1}{4} \left(1 - \sin^2 r_A \sin^2 r_B+ p \cos^2 r_A \cos^2 r_B - \sqrt{(\sin^2 r_A - \sin^2 r_B)^2 + 4p^2 \cos^2 r_A \cos^2 r_B}\right), \\
\lambda_2 &= \frac{1}{4} \left(1 + \sin^2 r_A + \sin^2 r_B + \sin^2 r_A \sin^2 r_B - p \cos^2 r_A \cos^2 r_B\right), \\
\lambda_3 &= \frac{1}{4} \left(1 - \sin^2 r_A \sin^2 r_B+ p \cos^2 r_A \cos^2 r_B + \sqrt{(\sin^2 r_A - \sin^2 r_B)^2 + 4p^2 \cos^2 r_A \cos^2 r_B}\right).
\end{align*}

The Bloch representation of quantum state $\rho_{A_{\text{I}} B_{\text{II}}}$: 
\begin{align*}
\rho_{A_{\text{I}} B_{\text{II}}} & = \frac{1}{4} \left( I \otimes I - \sin^2 r_A \sigma_3 \otimes I + \cos^2 r_B I \otimes \sigma_3 + p \cos r_A \sin r_B \sigma_1 \otimes \sigma_1 \right. \nonumber \\
& \quad \left. - p \cos r_A \sin r_B \sigma_2 \otimes \sigma_2 + (p \cos^2 r_A \sin^2 r_B - \sin^2 r_A \cos^2 r_B) \sigma_3 \otimes \sigma_3 \right), 
\end{align*}
the eigenvalues are as follows:
\begin{align*}
\lambda_0 &= \frac{1}{4} \left(1 - \sin^2 r_A - \cos^2 r_B + \sin^2 r_A \cos^2 r_B - p \cos^2 r_A \sin^2 r_B\right), \\
\lambda_1 &= \frac{1}{4} \left(1 - \sin^2 r_A \cos^2 r_B+ p \cos^2 r_A \sin^2 r_B - \sqrt{(\sin^2 r_A - \cos^2 r_B)^2 + 4p^2 \cos^2 r_A \sin^2 r_B}\right), \\
\lambda_2 &= \frac{1}{4} \left(1 + \sin^2 r_A + \cos^2 r_B + \sin^2 r_A \cos^2 r_B - p \cos^2 r_A \sin^2 r_B\right), \\
\lambda_3 &= \frac{1}{4} \left(1 - \sin^2 r_A \cos^2 r_B+ p \cos^2 r_A \sin^2 r_B + \sqrt{(\sin^2 r_A - \cos^2 r_B)^2 + 4p^2 \cos^2 r_A \sin^2 r_B}\right).
\end{align*}

The Bloch representation of quantum state $\rho_{A_{\text{II}} B_{\text{I}}}$: 
\begin{align*}
\rho_{A_{\text{II}} B_{\text{I}}} & = \frac{1}{4} \left( I \otimes I + \cos^2 r_A \sigma_3 \otimes I - \sin^2 r_B I \otimes \sigma_3 + p \sin r_A \cos r_B \sigma_1 \otimes \sigma_1 \right. \nonumber \\
& \quad \left. - p \sin r_A \cos r_B \sigma_2 \otimes \sigma_2 + (p \sin^2 r_A \cos^2 r_B - \cos^2 r_A \sin^2 r_B) \sigma_3 \otimes \sigma_3 \right), 
\end{align*}
the eigenvalues are as follows:
\begin{align*}
\lambda_0 &= \frac{1}{4} \left(1 - \cos^2 r_A - \sin^2 r_B + \cos^2 r_A \sin^2 r_B - p \sin^2 r_A \cos^2 r_B\right), \\
\lambda_1 &= \frac{1}{4} \left(1 - \cos^2 r_A \sin^2 r_B+ p \sin^2 r_A \cos^2 r_B - \sqrt{(\cos^2 r_A - \sin^2 r_B)^2 + 4p^2 \sin^2 r_A \cos^2 r_B}\right), \\
\lambda_2 &= \frac{1}{4} \left(1 + \cos^2 r_A + \sin^2 r_B + \cos^2 r_A \sin^2 r_B - p \sin^2 r_A \cos^2 r_B\right), \\
\lambda_3 &= \frac{1}{4} \left(1 - \cos^2 r_A \sin^2 r_B+ p \sin^2 r_A \cos^2 r_B + \sqrt{(\cos^2 r_A - \sin^2 r_B)^2 + 4p^2 \sin^2 r_A \cos^2 r_B}\right).
\end{align*}

The Bloch representation of quantum state $\rho_{A_{\text{II}} B_{\text{II}}}$: 
\begin{align*}
\rho_{A_{\text{II}} B_{\text{II}}} & = \frac{1}{4} \left( I \otimes I + \cos^2 r_A \sigma_3 \otimes I + \cos^2 r_B I \otimes \sigma_3 + p \sin r_A \sin r_B \sigma_1 \otimes \sigma_1 \right. \nonumber \\
& \quad \left. + p \sin r_A \sin r_B \sigma_2 \otimes \sigma_2 + (\cos^2 r_A \cos^2 r_B - p \sin^2 r_A \sin^2 r_B) \sigma_3 \otimes \sigma_3 \right), 
\end{align*}
the eigenvalues are as follows:
\begin{align*}
\lambda_0 &= \frac{1}{4} \left(1 - \cos^2 r_A - \cos^2 r_B + \cos^2 r_A \cos^2 r_B - p \sin^2 r_A \sin^2 r_B\right), \\
\lambda_1 &= \frac{1}{4} \left(1 - \cos^2 r_A \cos^2 r_B+ p \sin^2 r_A \sin^2 r_B - \sqrt{(\cos^2 r_A - \cos^2 r_B)^2 + 4p^2 \sin^2 r_A \sin^2 r_B}\right), \\
\lambda_2 &= \frac{1}{4} \left(1 + \cos^2 r_A + \cos^2 r_B + \cos^2 r_A \cos^2 r_B - p \sin^2 r_A \sin^2 r_B\right), \\
\lambda_3 &= \frac{1}{4} \left(1 - \cos^2 r_A \cos^2 r_B+ p \sin^2 r_A \sin^2 r_B + \sqrt{(\cos^2 r_A - \cos^2 r_B)^2 + 4p^2 \sin^2 r_A \sin^2 r_B}\right).
\end{align*}

\section{Appendix B}

Under the influence of phase flip channel, the Bloch representation and eigenvalues of quantum states $\rho^{pf}_{A_{\text{I}} B_{\text{I}}}$, $\rho^{pf}_{A_{\text{I}} B_{\text{II}}}$, $\rho^{pf}_{A_{\text{II}} B_{\text{I}}}$ and $\rho^{pf}_{A_{\text{II}} B_{\text{II}}}$ have the following expressions, respectively. 

The Bloch representation of quantum state $\rho^{pf}_{A_{\text{I}} B_{\text{I}}}$: 
\begin{align*}
\rho^{pf}_{A_{\text{I}} B_{\text{I}}} & = \frac{1}{4} \left( I \otimes I - \sin^2 r_A \sigma_3 \otimes I - \sin^2 r_B I \otimes \sigma_3 + (1-2k)^2p \cos r_A \cos r_B \sigma_1 \otimes \sigma_1 \right. \nonumber \\
& \quad \left. + (1-2k)^2p \cos r_A \cos r_B \sigma_2 \otimes \sigma_2 + (\sin^2 r_A \sin^2 r_B - p \cos^2 r_A \cos^2 r_B) \sigma_3 \otimes \sigma_3 \right), 
\end{align*}
the eigenvalues are as follows:
\begin{align*}
\lambda_0 &= \frac{1}{4} \left(1 - \sin^2 r_A - \sin^2 r_B + \sin^2 r_A \sin^2 r_B - p \cos^2 r_A \cos^2 r_B\right), \\
\lambda_1 &= \frac{1}{4} \left(1 - \sin^2 r_A \sin^2 r_B + p \cos^2 r_A \cos^2 r_B - \sqrt{(\sin^2 r_A - \sin^2 r_B)^2 + 4(1 - 2k)^4p^2 \cos^2 r_A \cos^2 r_B}\right),\\
\lambda_2 &= \frac{1}{4} \left(1 + \sin^2 r_A + \sin^2 r_B + \sin^2 r_A \sin^2 r_B - p \cos^2 r_A \cos^2 r_B\right), \\
\lambda_3 &= \frac{1}{4} \left(1 - \sin^2 r_A \sin^2 r_B + p \cos^2 r_A \cos^2 r_B + \sqrt{(\sin^2 r_A - \sin^2 r_B)^2 + 4(1 - 2k)^4p^2 \cos^2 r_A \cos^2 r_B}\right).
\end{align*}

The Bloch representation of quantum state $\rho^{pf}_{A_{\text{I}} B_{\text{II}}}$: 
\begin{align*}
\rho^{pf}_{A_{\text{I}} B_{\text{II}}} & = \frac{1}{4} \left( I \otimes I - \sin^2 r_A \sigma_3 \otimes I + \cos^2 r_B I \otimes \sigma_3 + (1-2k)^2p \cos r_A \sin r_B \sigma_1 \otimes \sigma_1 \right. \nonumber \\
& \quad \left. - (1-2k)^2p \cos r_A \sin r_B \sigma_2 \otimes \sigma_2 + (p \cos^2 r_A \sin^2 r_B - \sin^2 r_A \cos^2 r_B) \sigma_3 \otimes \sigma_3 \right), 
\end{align*}
the eigenvalues are as follows:
\begin{align*}
\lambda_0 &= \frac{1}{4} \left(1 - \sin^2 r_A - \cos^2 r_B + \sin^2 r_A \cos^2 r_B - p \cos^2 r_A \sin^2 r_B\right), \\
\lambda_1 &= \frac{1}{4} \left(1 - \sin^2 r_A \cos^2 r_B+ p \cos^2 r_A \sin^2 r_B - \sqrt{(\sin^2 r_A - \cos^2 r_B)^2 + 4(1 - 2k)^4p^2 \cos^2 r_A \sin^2 r_B}\right), \\
\lambda_2 &= \frac{1}{4} \left(1 + \sin^2 r_A + \cos^2 r_B + \sin^2 r_A \cos^2 r_B - p \cos^2 r_A \sin^2 r_B\right), \\
\lambda_3 &= \frac{1}{4} \left(1 - \sin^2 r_A \cos^2 r_B+ p \cos^2 r_A \sin^2 r_B + \sqrt{(\sin^2 r_A - \cos^2 r_B)^2 + 4(1 - 2k)^4p^2 \cos^2 r_A \sin^2 r_B}\right).
\end{align*}

The Bloch representation of quantum state $\rho^{pf}_{A_{\text{II}} B_{\text{I}}}$: 
\begin{align*}
\rho^{pf}_{A_{\text{II}} B_{\text{I}}} & = \frac{1}{4} \left( I \otimes I + \cos^2 r_A \sigma_3 \otimes I - \sin^2 r_B I \otimes \sigma_3 + (1-2k)^2p \sin r_A \cos r_B \sigma_1 \otimes \sigma_1 \right. \nonumber \\
& \quad \left. - (1-2k)^2p \sin r_A \cos r_B \sigma_2 \otimes \sigma_2 + (p \sin^2 r_A \cos^2 r_B - \cos^2 r_A \sin^2 r_B) \sigma_3 \otimes \sigma_3 \right), 
\end{align*}
the eigenvalues are as follows:
\begin{align*}
\lambda_0 &= \frac{1}{4} \left(1 - \cos^2 r_A - \sin^2 r_B + \cos^2 r_A \sin^2 r_B - p \sin^2 r_A \cos^2 r_B\right), \\
\lambda_1 &= \frac{1}{4} \left(1 - \cos^2 r_A \sin^2 r_B+ p \sin^2 r_A \cos^2 r_B - \sqrt{(\cos^2 r_A - \sin^2 r_B)^2 + 4(1 - 2k)^4p^2 \sin^2 r_A \cos^2 r_B}\right), \\
\lambda_2 &= \frac{1}{4} \left(1 + \cos^2 r_A + \sin^2 r_B + \cos^2 r_A \sin^2 r_B - p \sin^2 r_A \cos^2 r_B\right), \\
\lambda_3 &= \frac{1}{4} \left(1 - \cos^2 r_A \sin^2 r_B+ p \sin^2 r_A \cos^2 r_B + \sqrt{(\cos^2 r_A - \sin^2 r_B)^2 + 4(1 - 2k)^4p^2 \sin^2 r_A \cos^2 r_B}\right).
\end{align*}

The Bloch representation of quantum state $\rho^{pf}_{A_{\text{II}} B_{\text{II}}}$: 
\begin{align*}
\rho^{pf}_{A_{\text{II}} B_{\text{II}}} & = \frac{1}{4} \left( I \otimes I + \cos^2 r_A \sigma_3 \otimes I + \cos^2 r_B I \otimes \sigma_3 + (1-2k)^2p \sin r_A \sin r_B \sigma_1 \otimes \sigma_1 \right. \nonumber \\
& \quad \left. + (1-2k)^2p \sin r_A \sin r_B \sigma_2 \otimes \sigma_2 + (\cos^2 r_A \cos^2 r_B - p \sin^2 r_A \sin^2 r_B) \sigma_3 \otimes \sigma_3 \right), 
\end{align*}
the eigenvalues are as follows:
\begin{align*}
\lambda_0 &= \frac{1}{4} \left(1 - \cos^2 r_A - \cos^2 r_B + \cos^2 r_A \cos^2 r_B - p \sin^2 r_A \sin^2 r_B\right), \\
\lambda_1 &= \frac{1}{4} \left(1 - \cos^2 r_A \cos^2 r_B+ p \sin^2 r_A \sin^2 r_B - \sqrt{(\cos^2 r_A - \cos^2 r_B)^2 + 4(1 - 2k)^4p^2 \sin^2 r_A \sin^2 r_B}\right), \\
\lambda_2 &= \frac{1}{4} \left(1 + \cos^2 r_A + \cos^2 r_B + \cos^2 r_A \cos^2 r_B - p \sin^2 r_A \sin^2 r_B\right), \\
\lambda_3 &= \frac{1}{4} \left(1 - \cos^2 r_A \cos^2 r_B+ p \sin^2 r_A \sin^2 r_B + \sqrt{(\cos^2 r_A - \cos^2 r_B)^2 + 4(1 - 2k)^4p^2 \sin^2 r_A \sin^2 r_B}\right).
\end{align*}

\section{Appendix C}

Under the influence of bit flip channel, the Bloch representation and eigenvalues of quantum states $\rho^{bf}_{A_{\text{I}} B_{\text{I}}}$, $\rho^{bf}_{A_{\text{I}} B_{\text{II}}}$, $\rho^{bf}_{A_{\text{II}} B_{\text{I}}}$ and $\rho^{bf}_{A_{\text{II}} B_{\text{II}}}$ have the following expressions, respectively. 

The Bloch representation of quantum state $\rho^{bf}_{A_{\text{I}} B_{\text{I}}}$: 
\begin{align*}
\rho^{bf}_{A_{\text{I}} B_{\text{I}}} & = \frac{1}{4} \left( I \otimes I - (1-2k)\sin^2 r_A \sigma_3 \otimes I - (1-2k)\sin^2 r_B I \otimes \sigma_3 + p \cos r_A \cos r_B \sigma_1 \otimes \sigma_1 \right. \nonumber \\
& \quad \left. + (1-2k)^2p \cos r_A \cos r_B \sigma_2 \otimes \sigma_2 + (1-2k)^2(\sin^2 r_A \sin^2 r_B - p \cos^2 r_A \cos^2 r_B) \sigma_3 \otimes \sigma_3 \right), 
\end{align*}
the eigenvalues are as follows:
\begin{align*}
\lambda_0 &= \frac{1}{4} \left(1 + (1-2k)^2 \left(\sin^2 r_A \sin^2 r_B - p \cos^2 r_A \cos^2 r_B\right) \right. \nonumber \\
           & \quad \left. - \sqrt{(1-2k)^2\left(\sin^2 r_A + \sin^2 r_B\right)^2 + \left(1-(1 - 2k)^2\right)^2p^2 \cos^2 r_A \cos^2 r_B}\right), \\
\lambda_1 &= \frac{1}{4} \left(1 - (1-2k)^2 \left(\sin^2 r_A \sin^2 r_B - p \cos^2 r_A \cos^2 r_B\right) \right. \nonumber \\
           & \quad \left. - \sqrt{(1-2k)^2\left(\sin^2 r_A - \sin^2 r_B\right)^2 + \left(1+(1 - 2k)^2\right)^2p^2 \cos^2 r_A \cos^2 r_B}\right), \\[1ex]
\lambda_2 &= \frac{1}{4} \left(1 + (1-2k)^2 \left(\sin^2 r_A \sin^2 r_B - p \cos^2 r_A \cos^2 r_B\right) \right. \nonumber \\
           & \quad \left. + \sqrt{(1-2k)^2\left(\sin^2 r_A + \sin^2 r_B\right)^2 + \left(1-(1 - 2k)^2\right)^2p^2 \cos^2 r_A \cos^2 r_B}\right), \\
\lambda_3 &= \frac{1}{4} \left(1 - (1-2k)^2 \left(\sin^2 r_A \sin^2 r_B - p \cos^2 r_A \cos^2 r_B\right) \right. \nonumber \\
           & \quad \left. + \sqrt{(1-2k)^2\left(\sin^2 r_A - \sin^2 r_B\right)^2 + \left(1+(1 - 2k)^2\right)^2p^2 \cos^2 r_A \cos^2 r_B}\right).
\end{align*}

The Bloch representation of quantum state $\rho^{bf}_{A_{\text{I}} B_{\text{II}}}$: 
\begin{align*}
\rho^{bf}_{A_{\text{I}} B_{\text{II}}} & = \frac{1}{4} \left( I \otimes I - (1-2k)\sin^2 r_A \sigma_3 \otimes I + (1-2k)\cos^2 r_B I \otimes \sigma_3 + p \cos r_A \sin r_B \sigma_1 \otimes \sigma_1 \right. \nonumber \\
& \quad \left. - (1-2k)^2p \cos r_A \sin r_B \sigma_2 \otimes \sigma_2 + (1-2k)^2(2p \cos^2 r_A \sin^2 r_B - \sin^2 r_A \cos^2 r_B) \sigma_3 \otimes \sigma_3 \right), 
\end{align*}
the eigenvalues are as follows:
\begin{align*}
\lambda_0 &= \frac{1}{4} \left(1 - (1-2k)^2 \left(p \cos^2 r_A \sin^2 r_B - \sin^2 r_A \cos^2 r_B\right) \right. \nonumber \\
           & \quad \left. - \sqrt{(1-2k)^2\left(\sin^2 r_A + \cos^2 r_B\right)^2 + \left(1-(1 - 2k)^2\right)^2p^2 \cos^2 r_A \sin^2 r_B}\right), \\
\lambda_1 &= \frac{1}{4} \left(1 + (1-2k)^2 \left(p \cos^2 r_A \sin^2 r_B - \sin^2 r_A \cos^2 r_B\right) \right. \nonumber \\
           & \quad \left. - \sqrt{(1-2k)^2\left(\sin^2 r_A - \cos^2 r_B\right)^2 + \left(1+(1 - 2k)^2\right)^2p^2 \cos^2 r_A \sin^2 r_B}\right), \\
\lambda_2 &= \frac{1}{4} \left(1 - (1-2k)^2 \left(p \cos^2 r_A \sin^2 r_B - \sin^2 r_A \cos^2 r_B\right) \right. \nonumber \\
           & \quad \left. + \sqrt{(1-2k)^2\left(\sin^2 r_A + \cos^2 r_B\right)^2 + \left(1-(1 - 2k)^2\right)^2p^2 \cos^2 r_A \sin^2 r_B}\right), \\
\lambda_3 &= \frac{1}{4} \left(1 + (1-2k)^2 \left(p \cos^2 r_A \sin^2 r_B - \sin^2 r_A \cos^2 r_B\right) \right. \nonumber \\
           & \quad \left. + \sqrt{(1-2k)^2\left(\sin^2 r_A - \cos^2 r_B\right)^2 + \left(1+(1 - 2k)^2\right)^2p^2 \cos^2 r_A \sin^2 r_B}\right).
\end{align*}

The Bloch representation of quantum state $\rho^{bf}_{A_{\text{II}} B_{\text{I}}}$: 
\begin{align*}
\rho^{bf}_{A_{\text{II}} B_{\text{I}}} & = \frac{1}{4} \left( I \otimes I + (1-2k)\cos^2 r_A \sigma_3 \otimes I - (1-2k)\sin^2 r_B I \otimes \sigma_3 + p \sin r_A \cos r_B \sigma_1 \otimes \sigma_1 \right. \nonumber \\
& \quad \left. - (1-2k)^2p \sin r_A \cos r_B \sigma_2 \otimes \sigma_2 + (1-2k)^2(p \sin^2 r_A \cos^2 r_B - \cos^2 r_A \sin^2 r_B) \sigma_3 \otimes \sigma_3 \right), 
\end{align*}
the eigenvalues are as follows:
\begin{align*}
\lambda_0 &= \frac{1}{4} \left(1 - (1-2k)^2 \left(p \sin^2 r_A \cos^2 r_B - \cos^2 r_A \sin^2 r_B\right) \right. \nonumber \\
           & \quad \left. - \sqrt{(1-2k)^2\left(\cos^2 r_A + \sin^2 r_B\right)^2 + \left(1-(1 - 2k)^2\right)^2p^2 \sin^2 r_A \cos^2 r_B}\right), \\
\lambda_1 &= \frac{1}{4} \left(1 + (1-2k)^2 \left(p \sin^2 r_A \cos^2 r_B - \cos^2 r_A \sin^2 r_B\right) \right. \nonumber \\
           & \quad \left. - \sqrt{(1-2k)^2\left(\cos^2 r_A - \sin^2 r_B\right)^2 + \left(1+(1 - 2k)^2\right)^2p^2 \sin^2 r_A \cos^2 r_B}\right), \\
\lambda_2 &= \frac{1}{4} \left(1 - (1-2k)^2 \left(p \sin^2 r_A \cos^2 r_B - \cos^2 r_A \sin^2 r_B\right) \right. \nonumber \\
           & \quad \left. + \sqrt{(1-2k)^2\left(\cos^2 r_A + \sin^2 r_B\right)^2 + \left(1-(1 - 2k)^2\right)^2p^2 \sin^2 r_A \cos^2 r_B}\right), \\
\lambda_3 &= \frac{1}{4} \left(1 + (1-2k)^2 \left(p \sin^2 r_A \cos^2 r_B - \cos^2 r_A \sin^2 r_B\right) \right. \nonumber \\
           & \quad \left. + \sqrt{(1-2k)^2\left(\cos^2 r_A - \sin^2 r_B\right)^2 + \left(1+(1 - 2k)^2\right)^2p^2 \sin^2 r_A \cos^2 r_B}\right).
\end{align*}

The Bloch representation of quantum state $\rho^{bf}_{A_{\text{II}} B_{\text{II}}}$: 
\begin{align*}
\rho^{bf}_{A_{\text{II}} B_{\text{II}}} & = \frac{1}{4} \left( I \otimes I + (1-2k)\cos^2 r_A \sigma_3 \otimes I + (1-2k)\cos^2 r_B I \otimes \sigma_3 + p \sin r_A \sin r_B \sigma_1 \otimes \sigma_1 \right. \nonumber \\
& \quad \left. + (1-2k)^2p \sin r_A \sin r_B \sigma_2 \otimes \sigma_2 + (1-2k)^2(\cos^2 r_A \cos^2 r_B - p \sin^2 r_A \sin^2 r_B) \sigma_3 \otimes \sigma_3 \right), 
\end{align*}
the eigenvalues are as follows:
\begin{align*}
\lambda_0 &= \frac{1}{4} \left(1 + (1-2k)^2 \left(\cos^2 r_A \cos^2 r_B - p \sin^2 r_A \sin^2 r_B\right) \right. \nonumber \\
           & \quad \left. - \sqrt{(1-2k)^2\left(\cos^2 r_A + \cos^2 r_B\right)^2 + \left(1-(1 - 2k)^2\right)^2p^2 \sin^2 r_A \sin^2 r_B}\right), \\
\lambda_1 &= \frac{1}{4} \left(1 - (1-2k)^2 \left(\cos^2 r_A \cos^2 r_B - p \sin^2 r_A \sin^2 r_B\right) \right. \nonumber \\
           & \quad \left. - \sqrt{(1-2k)^2\left(\cos^2 r_A - \cos^2 r_B\right)^2 + \left(1+(1 - 2k)^2\right)^2p^2 \sin^2 r_A \sin^2 r_B}\right), \\
\lambda_2 &= \frac{1}{4} \left(1 + (1-2k)^2 \left(\cos^2 r_A \cos^2 r_B - p \sin^2 r_A \sin^2 r_B\right) \right. \nonumber \\
           & \quad \left. + \sqrt{(1-2k)^2\left(\cos^2 r_A + \cos^2 r_B\right)^2 + \left(1-(1 - 2k)^2\right)^2p^2 \sin^2 r_A \sin^2 r_B}\right), \\
\lambda_3 &= \frac{1}{4} \left(1 - (1-2k)^2 \left(\cos^2 r_A \cos^2 r_B - p \sin^2 r_A \sin^2 r_B\right) \right. \nonumber \\
           & \quad \left. + \sqrt{(1-2k)^2\left(\cos^2 r_A - \cos^2 r_B\right)^2 + \left(1+(1 - 2k)^2\right)^2p^2 \sin^2 r_A \sin^2 r_B}\right).
\end{align*}

\section{Appendix D}

Under the influence of depolarizing channel, the Bloch representation and eigenvalues of quantum states $\rho^{dep}_{A_{\text{I}} B_{\text{I}}}$, $\rho^{dep}_{A_{\text{I}} B_{\text{II}}}$, $\rho^{dep}_{A_{\text{II}} B_{\text{I}}}$ and $\rho^{dep}_{A_{\text{II}} B_{\text{II}}}$ have the following expressions, respectively. 

The Bloch representation of quantum state $\rho^{dep}_{A_{\text{I}} B_{\text{I}}}$: 
\begin{align*}
\rho^{dep}_{A_{\text{I}} B_{\text{I}}} & = \frac{1}{4} \left( I \otimes I - \Big(1-\frac{4k}{3}\Big)\sin^2 r_A \sigma_3 \otimes I - \Big(1-\frac{4k}{3}\Big)\sin^2 r_B I \otimes \sigma_3 + \Big(1-\frac{4k}{3}\Big)^2p \cos r_A \cos r_B \sigma_1 \otimes \sigma_1 \right. \nonumber \\
& \quad \left. + \Big(1-\frac{4k}{3}\Big)^2p \cos r_A \cos r_B \sigma_2 \otimes \sigma_2 + \Big(1-\frac{4k}{3}\Big)^2(\sin^2 r_A \sin^2 r_B - p \cos^2 r_A \cos^2 r_B) \sigma_3 \otimes \sigma_3 \right), 
\end{align*}
the eigenvalues are as follows:
\begin{align*}
\lambda_0 &= \frac{1}{4}\Big(1 - \Big(1-\frac{4k}{3}\Big)\sin^2 r_A
                - \Big(1-\frac{4k}{3}\Big)\sin^2 r_B \nonumber\\
           &\qquad\qquad\quad + \Big(1-\frac{4k}{3}\Big)^2\sin^2 r_A \sin^2 r_B
                - \Big(1-\frac{4k}{3}\Big)^2 p \cos^2 r_A \cos^2 r_B\Big), \\
\lambda_1 &= \frac{1}{4}\Big(1 - \Big(1-\frac{4k}{3}\Big)^2\sin^2 r_A \sin^2 r_B
                + \Big(1-\frac{4k}{3}\Big)^2 p \cos^2 r_A \cos^2 r_B \nonumber\\
           &\qquad\qquad\quad - \Big(1-\frac{4k}{3}\Big)
                \sqrt{(\sin^2 r_A - \sin^2 r_B)^2
                + 4\Big(1-\frac{4k}{3}\Big)^2 p^2 \cos^2 r_A \cos^2 r_B}\Big),\\
\lambda_2 &= \frac{1}{4}\Big(1 + \Big(1-\frac{4k}{3}\Big)\sin^2 r_A
                + \Big(1-\frac{4k}{3}\Big)\sin^2 r_B \nonumber\\
           &\qquad\qquad\quad + \Big(1-\frac{4k}{3}\Big)^2\sin^2 r_A \sin^2 r_B
                - \Big(1-\frac{4k}{3}\Big)^2 p \cos^2 r_A \cos^2 r_B\Big), \\
\lambda_3 &= \frac{1}{4}\Big(1 - \Big(1-\frac{4k}{3}\Big)^2\sin^2 r_A \sin^2 r_B
                + \Big(1-\frac{4k}{3}\Big)^2 p \cos^2 r_A \cos^2 r_B \nonumber\\
           &\qquad\qquad\quad + \Big(1-\frac{4k}{3}\Big)
                \sqrt{(\sin^2 r_A - \sin^2 r_B)^2
                + 4\Big(1-\frac{4k}{3}\Big)^2 p^2 \cos^2 r_A \cos^2 r_B}\Big).
\end{align*}

The Bloch representation of quantum state $\rho^{dep}_{A_{\text{I}} B_{\text{II}}}$: 
\begin{align*}
\rho^{dep}_{A_{\text{I}} B_{\text{II}}} & = \frac{1}{4} \left( I \otimes I - \Big(1-\frac{4k}{3}\Big)\sin^2 r_A \sigma_3 \otimes I + \Big(1-\frac{4k}{3}\Big)\cos^2 r_B I \otimes \sigma_3 + \Big(1-\frac{4k}{3}\Big)^2p \cos r_A \sin r_B \sigma_1 \otimes \sigma_1 \right. \nonumber \\
& \quad \left. - \Big(1-\frac{4k}{3}\Big)^2p \cos r_A \sin r_B \sigma_2 \otimes \sigma_2 + \Big(1-\frac{4k}{3}\Big)^2(2p \cos^2 r_A \sin^2 r_B - \sin^2 r_A \cos^2 r_B) \sigma_3 \otimes \sigma_3 \right), 
\end{align*}
the eigenvalues are as follows:
\begin{align*}
\lambda_0 &= \frac{1}{4}\Big(1
                - \Big(1-\frac{4k}{3}\Big)\sin^2 r_A
                - \Big(1-\frac{4k}{3}\Big)\cos^2 r_B \nonumber\\
           &\qquad\qquad\quad + \Big(1-\frac{4k}{3}\Big)^{2}\sin^2 r_A\cos^2 r_B
                - \Big(1-\frac{4k}{3}\Big)^{2}p\cos^2 r_A\sin^2 r_B\Big),\\
\lambda_1 &= \frac{1}{4}\Big(1
                - \Big(1-\frac{4k}{3}\Big)^{2}\sin^2 r_A\cos^2 r_B
                + \Big(1-\frac{4k}{3}\Big)^{2}p\cos^2 r_A\sin^2 r_B \nonumber\\
           &\qquad\qquad\quad - \Big(1-\frac{4k}{3}\Big)
                \sqrt{(\sin^2 r_A - \cos^2 r_B)^2
                + 4\Big(1-\frac{4k}{3}\Big)^{2}p^2\cos^2 r_A\sin^2 r_B}\Big),\\
\lambda_2 &= \frac{1}{4}\Big(1
                + \Big(1-\frac{4k}{3}\Big)\sin^2 r_A
                + \Big(1-\frac{4k}{3}\Big)\cos^2 r_B \nonumber\\
           &\qquad\qquad\quad + \Big(1-\frac{4k}{3}\Big)^{2}\sin^2 r_A\cos^2 r_B
                - \Big(1-\frac{4k}{3}\Big)^{2}p\cos^2 r_A\sin^2 r_B\Big),\\
\lambda_3 &= \frac{1}{4}\Big(1
                - \Big(1-\frac{4k}{3}\Big)^{2}\sin^2 r_A\cos^2 r_B
                + \Big(1-\frac{4k}{3}\Big)^{2}p\cos^2 r_A\sin^2 r_B \nonumber\\
           &\qquad\qquad\quad + \Big(1-\frac{4k}{3}\Big)
                \sqrt{(\sin^2 r_A - \cos^2 r_B)^2
                + 4\Big(1-\frac{4k}{3}\Big)^{2}p^2\cos^2 r_A\sin^2 r_B}\Big).
\end{align*}

The Bloch representation of quantum state $\rho^{dep}_{A_{\text{II}} B_{\text{I}}}$: 
\begin{align*}
\rho^{dep}_{A_{\text{II}} B_{\text{I}}} & = \frac{1}{4} \left( I \otimes I + \Big(1-\frac{4k}{3}\Big)\cos^2 r_A \sigma_3 \otimes I - \Big(1-\frac{4k}{3}\Big)\sin^2 r_B I \otimes \sigma_3 + \Big(1-\frac{4k}{3}\Big)^2p \sin r_A \cos r_B \sigma_1 \otimes \sigma_1 \right. \nonumber \\
& \quad \left. - \Big(1-\frac{4k}{3}\Big)^2p \sin r_A \cos r_B \sigma_2 \otimes \sigma_2 + \Big(1-\frac{4k}{3}\Big)^2(p \sin^2 r_A \cos^2 r_B - \cos^2 r_A \sin^2 r_B) \sigma_3 \otimes \sigma_3 \right), 
\end{align*}
the eigenvalues are as follows:
\begin{align*}
\lambda_0 &= \frac{1}{4}\Big(1
                - \Big(1-\frac{4k}{3}\Big)\cos^{2}r_A
                - \Big(1-\frac{4k}{3}\Big)\sin^{2}r_B \nonumber\\
           &\qquad\qquad\quad - \Big(1-\frac{4k}{3}\Big)^{2}
                \big(p\sin^{2}r_A\cos^{2}r_B
                - \cos^{2}r_A\sin^{2}r_B\big)\Big), \\[4pt]
\lambda_1 &= \frac{1}{4}\Big(1
                + \Big(1-\frac{4k}{3}\Big)^{2}
                \big(p\sin^{2}r_A\cos^{2}r_B
                - \cos^{2}r_A\sin^{2}r_B\big) \nonumber\\
           &\qquad\qquad\quad - \Big(1-\frac{4k}{3}\Big)
                \sqrt{(\cos^{2}r_A - \sin^{2}r_B)^2
                + 4\Big(1-\frac{4k}{3}\Big)^{2}p^{2}\sin^{2}r_A\cos^{2}r_B}\Big), \\[4pt]
\lambda_2 &= \frac{1}{4}\Big(1
                + \Big(1-\frac{4k}{3}\Big)\cos^{2}r_A
                + \Big(1-\frac{4k}{3}\Big)\sin^{2}r_B \nonumber\\
           &\qquad\qquad\quad - \Big(1-\frac{4k}{3}\Big)^{2}
                \big(p\sin^{2}r_A\cos^{2}r_B
                - \cos^{2}r_A\sin^{2}r_B\big)\Big), \\[4pt]
\lambda_3 &= \frac{1}{4}\Big(1
                + \Big(1-\frac{4k}{3}\Big)^{2}
                \big(p\sin^{2}r_A\cos^{2}r_B
                - \cos^{2}r_A\sin^{2}r_B\big) \nonumber\\
           &\qquad\qquad\quad + \Big(1-\frac{4k}{3}\Big)
                \sqrt{(\cos^{2}r_A - \sin^{2}r_B)^2
                + 4\Big(1-\frac{4k}{3}\Big)^{2}p^{2}\sin^{2}r_A\cos^{2}r_B}\Big).
\end{align*}

The Bloch representation of quantum state $\rho^{dep}_{A_{\text{II}} B_{\text{II}}}$: 
\begin{align*}
\rho^{dep}_{A_{\text{II}} B_{\text{II}}} & = \frac{1}{4} \left( I \otimes I + \Big(1-\frac{4k}{3}\Big)^2\cos^2 r_A \sigma_3 \otimes I + \Big(1-\frac{4k}{3}\Big)^2\cos^2 r_B I \otimes \sigma_3 + \Big(1-\frac{4k}{3}\Big)^2p \sin r_A \sin r_B \sigma_1 \otimes \sigma_1 \right. \nonumber \\
& \quad \left. + \Big(1-\frac{4k}{3}\Big)^2p \sin r_A \sin r_B \sigma_2 \otimes \sigma_2 + \Big(1-\frac{4k}{3}\Big)^2(\cos^2 r_A \cos^2 r_B - p \sin^2 r_A \sin^2 r_B) \sigma_3 \otimes \sigma_3 \right), 
\end{align*}
the eigenvalues are as follows:
\begin{align*}
\lambda_0 &= \frac{1}{4}\Big(1
                - \Big(1-\frac{4k}{3}\Big)\cos^2 r_A
                - \Big(1-\frac{4k}{3}\Big)\cos^2 r_B \nonumber\\
           &\qquad\qquad\quad + \Big(1-\frac{4k}{3}\Big)^{2}
                \big(\cos^{2}r_A\cos^{2}r_B
                - p\sin^{2}r_A\sin^{2}r_B\big)\Big),\\[4pt]
\lambda_1 &= \frac{1}{4}\Big(1
                - \Big(1-\frac{4k}{3}\Big)^{2}\cos^{2}r_A\cos^{2}r_B
                + \Big(1-\frac{4k}{3}\Big)^{2}p\sin^{2}r_A\sin^{2}r_B \nonumber\\
           &\qquad\qquad\quad - \Big(1-\frac{4k}{3}\Big)
                \sqrt{(\cos^{2}r_A - \cos^{2}r_B)^2
                + 4\Big(1-\frac{4k}{3}\Big)^{2}p^{2}\sin^{2}r_A\sin^{2}r_B}\Big),\\[4pt]
\lambda_2 &= \frac{1}{4}\Big(1
                + \Big(1-\frac{4k}{3}\Big)\cos^2 r_A
                + \Big(1-\frac{4k}{3}\Big)\cos^2 r_B \nonumber\\
           &\qquad\qquad\quad + \Big(1-\frac{4k}{3}\Big)^{2}
                \big(\cos^{2}r_A\cos^{2}r_B
                - p\sin^{2}r_A\sin^{2}r_B\big)\Big),\\[4pt]
\lambda_3 &= \frac{1}{4}\Big(1
                - \Big(1-\frac{4k}{3}\Big)^{2}\cos^{2}r_A\cos^{2}r_B
                + \Big(1-\frac{4k}{3}\Big)^{2}p\sin^{2}r_A\sin^{2}r_B \nonumber\\
           &\qquad\qquad\quad + \Big(1-\frac{4k}{3}\Big)
                \sqrt{(\cos^{2}r_A - \cos^{2}r_B)^2
                + 4\Big(1-\frac{4k}{3}\Big)^{2}p^{2}\sin^{2}r_A\sin^{2}r_B}\Big).
\end{align*}

\end{document}